\documentclass[prd,preprint,superscriptaddress,preprintnumbers,eqsecnum,showpacs,nofootinbib,nobibnotes]{revtex4}
\usepackage{amsfonts,amsmath,amssymb,bm,natbib}
\usepackage{graphicx} 
\usepackage{mathtools}
\usepackage{boondox-cal}
\usepackage[utf8]{inputenc}
\usepackage{slashed}

\newcommand{\be}{\begin{equation}}
\newcommand{\bea}{\begin{eqnarray}}
\newcommand{\ba}{\begin{align}}
\newcommand{\ee}{\end{equation}}
\newcommand{\eea}{\end{eqnarray}}
\newcommand{\ea}{\end{align}}

\def\1eq#1{Eq.~(\ref{#1})}

\def\2eqs#1#2{Eqs.~(\ref{#1}) and~(\ref{#2})}
\def\3eqs#1#2#3{Eqs.~(\ref{#1}),~(\ref{#2}) and~(\ref{#3})}
\def\noeq#1{(\ref{#1})}

\def\tbarc{\bar{\mathcal{c}}^*}

\def\fig#1{Fig.~\ref{#1}}

\def\s#1{{\scriptscriptstyle #1}}

\def\G{\Gamma}

\def\T{T_1}

\def\s{\mathcal{s}}

\def\hphi0{{\hat\phi}_0}

\begin{document}

\title{
Off-shell renormalization\\ in the presence of dimension~6 derivative operators.\\ 
I. General theory}

\date{February 23, 2019}

\author{D. Binosi}
\email{binosi@ectstar.eu}
\affiliation{European Centre for Theoretical Studies in Nuclear Physics
and Related Areas (ECT*) and Fondazione Bruno Kessler, Villa Tambosi, Strada delle Tabarelle 286, I-38123 Villazzano (TN), Italy}

\author{A. Quadri}
\email{andrea.quadri@mi.infn.it}
\affiliation{INFN, Sezione di Milano, via Celoria 16, I-20133 Milano, Italy}

\begin{abstract}
\noindent
The consistent recursive subtraction of UV divergences order by order in the loop expansion for spontaneously broken effective field theories with dimension-6 derivative operators is presented for an Abelian gauge group. We solve the Slavnov-Taylor identity to all orders in the loop expansion by homotopy techniques and a suitable choice of invariant field coordinates (named bleached variables) for the linearly realized gauge group. This allows one to disentangle the gauge-invariant contributions to off-shell 1-PI amplitudes from those associated with the gauge-fixing and (generalized) non-polynomial field redefinitions (that do appear already at one loop).
The tools presented can be easily generalized to the non-Abelian case.
%
\end{abstract}

\maketitle

\section{Introduction}

Whenever physics beyond the Standard Model (BSM) appears at an energy scale $\Lambda$ much higher than the electroweak scale $v$, it can be described, in the low energy regime, by an effective field theory (EFT). In this approach, physical operators of different mass dimension, compatible with the relevant symmetries of the theory, are arranged according to inverse powers of the scale $\Lambda$; and, in the so-called Standard Model Effective Field Theory (SMEFT), only operators up to dimension 6 are usually considered. 

As a consequence of the current lack of a direct evidence of BSM physics at the LHC~\cite{deFlorian:2016spz}, much effort in the recent literature has been poured into deriving the phenomenological implications of SMEFTs. If one is mainly interested in evaluating physical $S$-matrix elements in the classical or one-loop approximations (see~\cite{deFlorian:2016spz} for a recent review), only the knowledge of on-shell quantities is required, so that the classical equations of motion can be safely used  in order to discard operators that are equivalent on-shell. In addition, gauge-independent field reparametrizations which leave the $S$-matrix invariant can be carried out in order to cancel the highest possible number of operators, leaving eventually the basis of non-redundant operators classified in~\cite{Buchmuller:1985jz,Grzadkowski:2010es}. Perhaps, the most striking results obtained in this contest is a string of miraculous cancellations and regularities in the {\em brute force} one-loop evaluation of anomalous dimensions~\cite{Jenkins:2013zja,Jenkins:2013wua,Alonso:2013hga} which have been traced back to holomorphicity~\cite{Cheung:2015aba,Alonso:2014rga}, and/or remnants of embedding supersymmetry~\cite{Elias-Miro:2014eia}.
 
There are however a number of reasons to study the {\it off-shell} renormalization of EFTs in general, and SMEFTs in particular. The most obvious one is that these theories are supposed to be the low energy description of a yet unknown UV complete theory; indeed, matching physical gauge-invariant quantities with the corresponding UV theory predictions at the EFT cutoff scale should in principle be sufficient to  obtain an EFT that properly reproduces the UV theory predictions at lower energies. However, this task can be accomplished iff the low energy EFT respects locality, that is: UV divergences remain local to all-order in perturbation theory; overlapping divergences are appropriately subtracted according to Bobolyubov $R$-operation~\cite{Bogoliubov:1957gp,Hepp:1966eg,Bogolyubov:1980nc}, or, equivalently, Zimmermann forest formula~\cite{Zimmermann:1969jj}; and, finally, the theory's defining functional identities such as the Slavnov-Taylor (ST) identity are preserved to all-order. In particular, the latter identity is essential to ensure the cancellations of unphysical ghost modes, and thus unitarity\footnote{This is the notion of {\it physical} unitarity, not to be confused with the fulfilment of the Froissart bound for the asymptotic energy behavior of physical quantities~\cite{Froissart:1961ux} which is not guaranteed to happen for UV incomplete theories, like EFTs.}. The locality requirement can be satisfied only when a proper renormalization of off-shell amplitudes is performed, thus leading to the appropriate generalized field redefinitions. While the UV divergent parts of the latter will be uniquely fixed order by order in the perturbative expansion, their finite parts (which will not affect physical observables) can be arbitrarily chosen, provided that they preserve the ST identity. Also, Renormalization-Group Equations (RGEs) for EFTs require, beyond the one-loop order, the consistent off-shell renormalization of the theory~\cite{Buchler:2003vwr}; indeed, higher orders RGEs are needed in order to compute subleading logarithmic divergences of physical observables; hence, off-shell renormalization cannot be avoided in order to extract the full physical information from EFTs~\cite{Weinberg:1978kz,Gasser:1983yg,Buchler:2003vwr}.

It turns out that exploiting the (gauge) symmetries of EFTs has the potential to lead to a deeper understanding of the inner workings of such an off-shell renormalization procedure. These symmetries can be treated in a mathematically consistent way in the so-called Batalin-Vilkovisky (BV) formalism~\cite{Gomis:1994he,Batalin:1984jr,Batalin:1981jr} where: Gauge symmetry is lifted to Becchi-Rouet-Stora-Tyutin (BRST) symmetry after gauge-fixing the classical action; for each field an external source, known as the antifield, is coupled to the BRST transformation of the field; and, finally, BRST invariance is encoded in a functional identity known as the BV master equation, or ST identity, which, for anomaly-free theories, holds to all loop orders~\cite{Slavnov:1972fg,Taylor:1971ff,Gomis:1994he,Batalin:1984jr,Batalin:1981jr}. It has been proven a long time ago ~\cite{Gomis:1995jp} that for anomaly-free EFTs UV divergences can be consistently removed while respecting the BV master equation; this is achieved through an appropriate choice of all possible gauge-invariant operators, supplemented by a canonical redefinition of the fields and antifields of the theory that generalize to the non power-counting renormalizable case the familiar linear field redefinitions of the renormalizable theories.

Obviously such field redefinitions cannot be chosen arbitrarily, since they are constrained by the fulfilment of the BV master equation. More specifically, their form is fixed order by order in the loop expansion by the UV divergences of amplitudes involving antifields. As we will show, these restrictions are rather strong: If a field redefinition is carried out at the quantum level without taking them into account, then the locality property of higher order counterterms is lost; plainly, the divergences cannot be anymore consistently removed satisfying the locality requirement described above.

The locality properties of 1-PI Green's functions are encoded in the so-called Quantum Action Principle (QAP)~\cite{Lam:1972mb,Breitenlohner:1977hr,Zimmermann:1969jj,Zimmermann:1972te} and intimately related to the topological loop expansion; therefore, the loop order is the appropriate parameter expansion in the symmetric perturbative treatment based on the fulfillment of the BV master equation {\em \`a la} Weinberg\footnote{We notice that this is not necessarily the ordering of the size of the contributions of higher-dimensional operators to physical quantities (for a recent discussion of the different problem of devising the power-counting in momenta as a tool for predicting the energy dependence of physical quantities in the EFT framework see \cite{Gavela:2016bzc}).}.  The set of consistency conditions dictated by the BV bracket can be fully solved in a rather efficient way by combining a novel tool based on the idea of~{\em bleaching}~\cite{Bettinelli:2007tq,Bettinelli:2007cy,Bettinelli:2008ey,Bettinelli:2008qn,Bettinelli:2009wu,Bettinelli:2014msa,Binosi:2012cz} in the context of an EFT model with a linearly realized gauge symmetry and homotopy techniques in order to control the antifield-dependent sector of the theory. It turns out that the full dependence on the Goldstone fields can be completely determined in a purely algebraic fashion order by order in the loop expansion.

This remarkable fact amounts to the statement that for spontaneously broken gauge theories the ST identity can be explicitly solved. This in turn provides a very powerful tool to disentangle the relations between the higher dimension operators, induced by the ST identity, that are difficult to manage via the usual invariant expansion, due to the large number of operators involved in an EFTs where no on-shell equivalence between operators is enforced.

EFTs in the presence of derivative interactions usually
exhibit some sort of resummations of the insertions 
of higher dimensional operators, so that not all the
amplitudes that are  UV divergent are indeed independent.
The formalism used in this paper makes  use of a particular choice of field coordinates, the so-called $X$-formalism,
that is well-suited in order to deal with the decomposition
of amplitudes in order to keep track of the underlying
relations between the UV divergent coefficients.
In Appendix~\ref{app:toy} we describe a scalar toy model where we illustrate the advantages of the $X$-formalism in the classification of the independent UV divergences and their resummation.
The present paper is devoted to the study of the off-shell renormalization of an Abelian Higgs-Kibble model\footnote{The formalism can be extended directly to the non-Abelian case, as we will discuss in detail.} supplemented by the dimension 6 operator $\phi^\dagger \phi (D_\mu \phi)^\dagger D^\mu\phi$. The latter has been chosen as a non-trivial test of the formalism we are going to develop since it generates interaction vertices with two derivatives, thus leading to a maximal violation of the power-counting already at one loop (an infinite number of UV divergent amplitudes in fact exists at the one-loop level as a consequence of the presence of the 2-derivative interactions). Explicit computations are carried out at the one loop level, while the algebraic tools presented can be applied to all orders in the loop expansion.

The paper is organized as follows. In Sect.~\ref{sec:model} we set our notations and reformulate the Higgs-Kibble model supplemented by the maximally power counting violating dimension 6 operator $\phi^\dagger \phi (D_\mu \phi)^\dagger D^\mu \phi$ in the $X$-formalism (i.e. one uses as field coordinate to describe the physical Higgs scalar the gauge invariant bilinear $X_2 \sim \phi^\dagger \phi - \frac{v^2}{2}$, where $\langle \phi \rangle = \frac{v}{\sqrt{2}}$ is the vacuum expectation value). The identities obeyed by the classical vertex functional, encoding the symmetries of the theory, are then derived in Sect.~\ref{sec:fids}. In Sect.~\ref{sec:pc} we show that in the $X$-formulation a residual power counting is present for certain amplitudes (called ancestor amplitudes), which will be shown to be enough to generate order by order all the divergent amplitudes in the original formulation. Sect.~\ref{sec:sol.and.stab} contains the central results of this paper, as we proceed to study the solution and stability of all the functional identities for the complete vertex functional, {\it i.e.}, without locality restrictions. This is achieved by combining the bleaching of the field coordinates (an operatorial-valued finite gauge transformation leading to invariant variables) with homotopy techniques designed to deal with the non gauge-invariant contributions to the 1-PI amplitudes. The latter are controlled by the antifield-depedendent 1-PI Green's functions, encoding the remnant of the gauge-fixing and the generalized field redefinitions. We stress that this tool allows one to recursively obtain the solution to the ST identity to all orders in the loop expansion without any locality restriction. As an example we obtain the one-loop two point Goldstone and mixed gauge-Goldstone amplitudes and check that they verify the conditions imposed by the ST identity, as expected. In passing we will also identify and describe the procedure of the recursive subtraction of the divergences for off-shell 1-PI amplitudes, order by order in the loop expansion. In Sect.~\ref{sec:mapping} we consider some applications of the formalism by deriving several identities for the Green's functions of the model in the standard ordinary $\phi$-representation. In particular, we show that there are indeed non-polynomial field redefinitions that have to be taken into account. Then we move to the study of the two-point Higgs Green's function. We exploit the mapping from the $X$-theory to the standard formalism in order to separate the amplitude contributions according to their gauge transformation properties; in particular, we isolate the effects of field redefinitions and spot the genuinely new physical operator giving rise to the four-momentum contribution to the one loop Higgs two-point function. Lastly we describe the procedure for the extraction of the coefficients of higher dimensional operators and as a non-trivial example we study the renormalization of the radiatively generated operator $F_{\mu\nu}^2 \Big (\phi^\dagger \phi - \frac{v^2}{2} \Big )$. The application to other dimension 6 operators at one loop order is presented in a companion paper, devoted to the full one-loop  renormalization of such operators. The extension to non-Abelian gauge theories is addressed in Sect.~\ref{sec:nonabelian}. Conclusions are finally presented in Section~\ref{sec:conclusions}, followed by four appendices collecting results used throughout the presentation.

\section{\label{sec:model}The Model and Its Symmetries}

As has been shown in~\cite{Quadri:2016wwl} the spontaneous symmetry breaking (SSB) mechanism can be reformulated using as a dynamical variable the gauge invariant combination $\phi^\dagger \phi - \frac{v^2}{2}$,  where $\phi = \frac{1}{\sqrt{2}} (\sigma + v + i \chi)$ is the complex scalar field, $\chi$ the Goldstone, and $\sigma$ the Higgs scalar. We denote by $X_2$ the field coordinate for such a  gauge invariant combination. Then, using the same notation  as in~\cite{Binosi:2017ubk}, the action of the Abelian Higgs-Kibble model supplemented by the dim.6 operator $X_2 (D_\mu \phi)^\dagger D^\mu \phi \sim (\phi^\dagger \phi - \frac{v^2}{2}) (D_\mu \phi)^\dagger D^\mu \phi$ can be written as
\begin{align}
	S  = &  \int \!\mathrm{d}^4x \, \Big [ -\frac{1}{4} F^{\mu\nu} F_{\mu\nu} + (D^\mu \phi)^\dagger (D_\mu \phi) - \frac{M^2-m^2}{2} X_2^2 - \frac{m^2}{2v^2} \Big ( \phi^\dagger \phi - \frac{v^2}{2} \Big )^2 \nonumber \\
	& - \bar c (\square + m^2) c + \frac{1}{v} (X_1 + X_2) (\square + m^2) \Big ( \phi^\dagger \phi - \frac{v^2}{2} - v X_2 \Big ) \nonumber \\
	& + \frac{g}{\Lambda} X_2 (D^\mu \phi)^\dagger (D_\mu \phi)  + \T(D^\mu \phi)^\dagger (D_\mu \phi) \Big ].
	\label{cl.g.i.act}
\end{align}
In the formula above, we denote by $A_\mu$ the Abelian gauge connection; $\Lambda$ is then the new physics scale, and $g$ represents the dimensionless coupling constant of the dimension-6 operator. Moreover $T_1$ is an external source required in the formulation of the functional identities controlling the Algebraic Renormalization of the theory, as we will discuss shortly. The field $X_1$ is a Lagrange multiplier: by going on-shell one obtains
\footnote{On general grounds the $X_1$-equation of motion $\frac{1}{v} (\square + m^2) \Big ( \phi^\dagger \phi - \frac{v^2}{2} - v X_2 \Big ) =0$
would imply
$X_2=\frac{1}{v} \Big ( \phi^\dagger \phi - \frac{v^2}{2} \Big ) + \eta$, $\eta$ being a field fulfilling the Klein-Gordon equation
$(\square + m^2) \eta =0 $. In Sect.~\ref{sec:fids} we will show that in perturbation theory the correlators of $\eta$ with any gauge-invariant operators are zero, as a consequence of the $X_1$-equation; therefore it is consistent to set $\eta=0$.}
in fact the tree-level constraint
\begin{align}
	\frac{\delta S}{\delta X_1}
	&=0& 
	&
	\Longrightarrow & X_2 = \frac{1}{v} \Big ( \phi^\dagger \phi - \frac{v^2}{2} \Big ),&
	\label{X1EOM}
\end{align}
and the dimension-6 operator in the last line of Eq. (\ref{cl.g.i.act}) takes the more familiar form $\frac{g}{v \Lambda} \Big ( \phi^\dagger \phi - \frac{v^2}{2} \Big )(D^\mu \phi)^\dagger (D_\mu \phi)$. 

The model describes a vector meson of mass $M_A = e v$ and a physical scalar excitation $X_2$ of mass $M$; as already said, $\chi$ is the unphysical Goldstone boson associated with SSB, whereas the field $\sigma$ can be traded for  in favour of the unphysical mass eigenstate combination $\sigma'=\sigma-X_1-X_2$.  Both $\sigma'$ and $X_1$ have mass $m^2$ and their propagator differ by a sign, so that they cancel against each other in amplitudes of gauge-invariant operators~\cite{Quadri:2016wwl}. This cancellation can be seen as a consequence of an additional BRST symmetry of the theory, which reads
\begin{align}
	\s X_1 = v c; \, \quad \s \phi = \s X_2 = \s c  = 0; \quad \s \bar c = \phi^\dagger\phi - \frac{v^2}{2} - v X_2, 
	\label{u1.brst}
\end{align}
and guarantees that no further physical degrees of freedom  are introduced in addition to the gauge field and the physical scalar~\cite{Quadri:2006hr,Quadri:2016wwl}. We call this BRST symmetry {\em constraint} BRST symmetry as opposed to the {\it gauge group} BRST symmetry of the classical action after gauge-fixing.

We remark that the propagators of the field $X=X_1+X_2$ fall off as $1/p^4$ for large momenta~\cite{Quadri:2016wwl}
\begin{align}
    \Delta_{XX} = \Delta_{X\sigma} =\Delta_{X_1X_1}+\Delta_{X_2X_2} = \frac{i (M^2-m^2)}{(p^2-m^2)(p^2-M^2)}.
\end{align}
Since at $g=0$ the potentially power-counting violating interaction vertices of dimension~$5$ in the second line of Eq.(\ref{cl.g.i.act}) only involve the field $X$, the improved UV behaviour of the $X$-propagators ensures that the model is still power-counting renormalizable. Once the dimension-6 operator in the last line of Eq.~(\ref{cl.g.i.act})  is switched on, vertices involving the single $X_2$ field (and not the combination $X$) appear, leading to the violation of power-counting renormalizability by contributions proportional to $g$. The $X$-theory formalism requires that at $X_2=0$ the model reduces to the power-counting renormalizable theory.

As already mentioned, in addition to the constraint BRST invariance, the classical action is also invariant under the BRST symmetry  obtained by replacing the infinitesimal gauge parameter with the ghost $\omega$, so that
\begin{eqnarray}
	s A_\mu = \partial_\mu \omega \, ;  \qquad s \omega = 0 \, ; \qquad  s \bar{\omega} = b \, ; \qquad s b =0 \, ; \quad s \phi = i e \omega \phi.  
\end{eqnarray}
Both $\s$ and $s$ are nilpotent and anticommute.

The action~(\ref{cl.g.i.act}) needs to be gauge fixed. We choose a $R_\xi$-gauge-fixing, which is carried out {\em \`a la} BRST by introducing a pair of antighost and ghost fields $\bar{\omega},\omega$ and a Nakanishi-Lautrup (NL) multiplier field $b$:
\begin{align}
	S_\mathrm{gf} = & \int \!\mathrm{d}^4x \, s \Big[  \bar{\omega} \Big( \frac{1}{2\xi} b - \partial A - \frac{e v}{\xi} \chi\Big) \Big]   \nonumber \\
	=  & \int \!\mathrm{d}^4x \, \Big [ \frac{b^2}{2\xi} -  b \Big ( \partial A + \frac{e v}{\xi} \chi \Big) + \bar{\omega} \Big( \square \omega + \frac{e^2 v}{\xi} (\sigma + v) \omega \Big)  \Big ] .
	\label{g.f.}  
\end{align}
Explicit computations will be carried out in this paper in the Feynman gauge $\xi=1$.

Finally we introduce a set of external sources (antifields) coupled to the non-linear BRST transformations of the fields (for linear BRST transformations use of the antifields can be avoided~\cite{Piguet:1995er} since these transformations do not get an independent renormalization with respect to the quantized fields):
\begin{align}
	S_\mathrm{ext} = & \int \!\mathrm{d}^4x \, \Big [ \bar c^* \Big ( \phi^\dagger \phi - \frac{v^2}{2} - v X_2 \Big ) + \sigma^* (-e \omega \chi) + \chi^* e \omega (\sigma + v) \Big ].
	\label{ext.src}
\end{align}

The full tree-level vertex functional is, finally,
\begin{align}
	\G^{(0)} & = S + S_\mathrm{gf}  + S_\mathrm{ext} \nonumber \\
	& =   \int \!\mathrm{d}^4x \, \Big [ -\frac{1}{4} F^{\mu\nu} F_{\mu\nu} + (D^\mu \phi)^\dagger (D_\mu \phi) - \frac{M^2-m^2}{2} X_2^2 - \frac{m^2}{2v^2} \Big ( \phi^\dagger \phi - \frac{v^2}{2} \Big )^2 \nonumber \\
	& - \bar c (\square + m^2) c + \frac{1}{v} (X_1 + X_2) (\square + m^2) \Big ( \phi^\dagger \phi - \frac{v^2}{2} - v X_2 \Big ) \nonumber \\
	&  + \frac{g}{\Lambda} X_2 (D^\mu \phi)^\dagger (D_\mu \phi)  + \T(D^\mu \phi)^\dagger (D_\mu \phi) \nonumber \\
	& + \frac{b^2}{2\xi} -  b \Big ( \partial A + \frac{e v}{\xi} \chi \Big ) + \bar{\omega}\Big ( \square \omega + \frac{e^2 v}{\xi} (\sigma + v) \omega\Big ) \nonumber \\
	&  + \bar c^* \Big ( \phi^\dagger \phi - \frac{v^2}{2} - v X_2 \Big ) + \sigma^* (-e \omega \chi) + \chi^* e \omega (\sigma + v) \Big ].
	\label{tree.level}
\end{align}
The propagators of all fields are summarized in Appendix~\ref{app.prop}, whereas the ghost number is~$+1$ for $c$ and $\omega$, $-1$ for $\bar c$, $\bar \omega$, $\sigma^*$, $\chi^*$, and $0$ for all the remaining fields and external sources (obviously the vertex functional is ghost neutral). Finally, under charge conjugation $A_\mu, \chi, b$, $\omega,\bar\omega$ and the antifield $\chi^*$ are $C$-odd, while all other fields and external sources are $C$-even, as the effective action is.

Notice, finally, that this approach, with the set of external sources introduced above, is applicable only to models where the condition that higher-dimensional operators vanish at $X_2=0$ can be imposed; as it stands, it cannot handle a theory with, {\it e.g.}, the tree-level insertion of the operator $(D^\mu D^2 \phi)^\dagger D_\mu D^2\phi$. On the other hand, the generalization to this case is straightforward and briefly discussed in Appendix~\ref{app:Gauss} where the off-shell equivalence between the $X$- and the target theory is also proven.

\section{\label{sec:fids}Functional identities}

The tree-level functional~(\ref{tree.level}) obeys a set of functional identities which we list in the following
\begin{enumerate}
\item The $b$-equation:
\begin{eqnarray}
	\frac{\delta \G^{(0)}}{\delta b} = \frac{b}{\xi} - \partial A - \frac{e v}{\xi} \chi ;
	\label{b.eq}
\end{eqnarray}
\item The antighost equation:
\begin{eqnarray}
	\frac{\delta \G^{(0)}}{\delta \bar \omega} = \square \omega + \frac{e v}{\xi} \frac{\delta \G^{(0)}}{\delta \chi^*}  ;
	\label{antigh.eq}
\end{eqnarray}
\item The $X_1$-equation:
\begin{eqnarray}
	\frac{\delta \G^{(0)}}{\delta X_1} = \frac{1}{v} (\square + m^2) \frac{\delta \G^{(0)}}{\delta \bar c^*} ;
	\label{X1.eq}
\end{eqnarray}
\item The $X_2$-equation:
\begin{eqnarray}
	\frac{\delta \G^{(0)}}{\delta X_2} =  \frac{1}{v} (\square + m^2) \frac{\delta \G^{(0)}}{\delta \bar c^*} + \frac{g}{\Lambda} \frac{\delta \G^{(0)}}{\delta T_1} - (\square + m^2)X_1 - (\square + M^2) X_2 - v \bar c^*  ;
	\label{X2.eq}
\end{eqnarray}
\item The constraint ghost and antighost equations:
\begin{eqnarray}
	\frac{\delta \G^{(0)}}{\delta c} = (\square + m^2) \bar c; \qquad
	\frac{\delta \G^{(0)}}{\delta \bar c} = - (\square + m^2) c;
	\label{constr.antigh.eq}
\end{eqnarray}
\item The BV master equation / ST identity:
\begin{align}
	& {\cal S}(\G^{(0)})  = \int \mathrm{d}^4x \, \Big [ 
	\partial_\mu \omega \frac{\delta \G^{(0)}}{\delta A_\mu} + \frac{\delta \G^{(0)}}{\delta \sigma^*} \frac{\delta \G^{(0)}}{\delta \sigma}  + \frac{\delta \G^{(0)}}{\delta \chi^*} \frac{\delta \G^{(0)}}{\delta \chi} 
	+ b \frac{\delta \G^{(0)}}{\delta \bar \omega} \Big ] = 0. 
	\label{sti} 
\end{align}
\end{enumerate}
Notice that the ST identity  associated with the constraint BRST symmetry is not an independent equation, since by using the second of Eqs.(\ref{constr.antigh.eq}) one gets
\begin{align}
& {\cal S}_{\scriptscriptstyle{C}}(\G^{(0)}) \equiv \int \!\mathrm{d}^4 x \, \Big [ v c \frac{\delta \G^{(0)}}{\delta X_1} 
 + \frac{\delta \G^{(0)}}{\delta \bar c^*}\frac{\delta \G^{(0)}}{\delta \bar c} \Big ] = 
 \int \!\mathrm{d}^4 x \, \Big [ v c \frac{\delta \G^{(0)}}{\delta X_1} 
 -(\square + m^2) c \frac{\delta \G^{(0)}}{\delta \bar c^*} \Big ] = 0,
 \label{sti.c} 
\end{align}
which is the same as the $X_1$-equation~\noeq{X1.eq} since the constraint ghost $c$ is free.

Concerning the $X_{1,2}$ equations, it is instructive to introduce the generating functional for the connected amplitudes $W$ by taking the usual Legendre transform of $\G$ w.r.t. the fields~$\Phi$
\begin{align}
    W &= \G + \int \!\mathrm{d}^4 x \,  J \Phi; & \frac{\delta W}{\delta J} = \Phi;&  &\frac{\delta W}{\delta \zeta} = \frac{\delta \G}{\delta \zeta},
\end{align}
where $J$ represents a collective notation for the sources of the quantized fields $\Phi$, whereas $\zeta$ is a collective notation for the other external sources in $\G$.

Then at the connected level \2eqs{X1.eq}{X2.eq}  become
\begin{subequations}
	\begin{align}
    	& -J_{X_1} = \frac{1}{v}(\square + m^2) \frac{\delta W}{\delta \bar c^*} , \\
    	& -J_{X_2} = \frac{1}{v}(\square + m^2) \frac{\delta W}{\delta \bar c^*} + \frac{g}{\Lambda} \frac{\delta W}{\delta T_1} - (\square + m^2)\frac{\delta W}{\delta J_{X_1}} - (\square + M^2) \frac{\delta W}{\delta J_{X_2}} - v \bar c^* .
	\end{align}
\end{subequations}
By differentiating the first of the above equations w.r.t. any source $\zeta_i(y)$ and then going on shell by setting $J=\zeta=0$ we obtain
\begin{align}
    \frac{1}{v} (\square_x + m^2) W_{\zeta_i \bar c^*}(y,x) = 0 \, .
\label{insertion.barc}
\end{align}
The above equation implies
in perturbation theory that
\begin{align}
    W_{\zeta_i \bar c^*}(y,x) = 0. 
    \label{two.ins.barc}
\end{align}
This condition has an intuitive meaning: it simply states that the insertion of the composite operator $\phi^\dagger \phi - \frac{v^2}{2} - v X_2$, coupled to $\bar c^*$, vanishes when going on-shell with the source of the Lagrange multiplier $X_1$, enforcing the constraint. Since \1eq{insertion.barc} is valid to all orders in the loop expansion, it implies that the constraint is radiatively stable, as a consequence of the~$X_1$-equation. Explicit one-loop consistency checks of \1eq{two.ins.barc} are given in Appendix~\ref{app:conn_amps}.

We remark that Eq.(\ref{two.ins.barc}) 
implies  the absence of contributions from zero modes of the Klein-Gordon operator 
entering into the $X_1$-equation of motion, so that the condition
\begin{align}
    \frac{\delta S}{\delta X_1} = 
    \frac{1}{v} (\square + m^2)
    \Big ( \phi^\dagger \phi - \frac{v^2}{2} - v X_2 \Big ) = 0, 
\end{align}
indeed yields $X_2 = \frac{1}{v}
\Big ( \phi^\dagger \phi - \frac{v^2}{2}\Big ).$

\section{\label{sec:pc}Power counting}

From the previous section it clearly appears that there are two class of fields in our model: those whose Green's functions are uniquely fixed by the functional identities introduced above (namely $b,\bar c, c, \bar \omega, X_{1,2}$ and the Goldstone field $\chi$, which is controlled in a non-trivial way, as we will show, by the ST identity); and the gauge field $A_\mu$, the Higgs scalar $\sigma$ and the external sources, for which the functional identities are not effective. This allows in turn to introduce two class of amplitudes:  {\em ancestor amplitudes} involving the fields $A_\mu, \sigma$ and/or external sources; and {\em descendant amplitudes} involving at least one insertion of the remaining fields.

It turns out that when formulated in terms of the $X$ fields, the model at hand exhibits power-counting bounds which will limit to a finite number the set of UV divergent ancestor and $\chi$-amplitudes; obviously, since we are dealing with an EFT, this number will be increasing with the loop order. The only exception are the amplitudes involving the $T_1$ source, for which an unbounded dependence arises from the insertion of the vertex $\sim T_1 (\partial_\mu \sigma \partial^\mu \sigma + \partial_\mu \chi \partial^\mu \chi )$ on the $\sigma'$-, $X_1$-, $X_2$- and $\chi$-propagators; luckily, however, resummation of these amplitudes is possible in all cases, as we will explicitly show later for the one-loop UV divergent antifield-ghost amplitude.

\begin{figure}
    \centering
    \includegraphics[scale=.75]{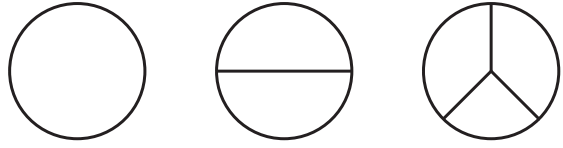}
    \caption{Most UV divergent one-, two- and three-loop vacuum topologies}
    \label{fig.1}
\end{figure}

Consider then the vacuum topologies. The derivative-interaction vertices in~\1eq{tree.level} are trilinear and thus the most UV divergent $n$-loop vacuum topologies (with $n>1$) are those with the maximum number of trilinear vertices (some sample topologies are represented in \fig{fig.1}).
For $n>1$ the UV degree of divergence $\delta_{n}$ of these diagrams can be obtained as follows. Each interaction vertex contributes two powers of the momenta. There are $V=2(n-1)$ of such vertices. Each propagator contributes two inverse powers of the momenta. According to Euler's relation the number of internal lines is $i=V + n -1 = 3(n-1)$. In $D$ dimensions each loop integration contributes a factor of $D$ to $\delta_n$, so that
\begin{align}
    \delta_n = n D + 4 (n-1) - 6 (n-1) = n(D-2) + 2.
\end{align}
(For the case $n=1$ the formula gives $\delta_1=D$, i.e. it accounts for the loop integration).

Then one can start inserting external legs on these topologies. Let us first discuss the $T_1=0$ sector. The insertion of one interaction vertex changes the degree of divergence by: increasing it with the powers of momenta of the vertex on the internal lines of the diagram; reducing it by 2, since a new propagator is inserted. Thus a vertex with no derivatives reduces the degree of divergence by $2$, a one-derivative (on the internal line) vertex reduces it by $1$ and a two-derivative (on internal lines) vertex does not change the degree of divergence at all. Thus, the most divergent $n$-th loop amplitudes are obtained by maximizing the number of insertions of vertices with two and one derivatives.

Let us enumerate these vertices. Since we are only interested in ancestor amplitudes, the dimension 6 operator $X_2 (D^\mu \phi)^\dagger D_\mu \phi$ only generates vertices with one derivative on internal lines ($X_2$ must in fact be an internal line). In addition
we are only interested in the vertex types $X_2 \partial^\mu \sigma \partial_\mu \sigma$
and $X_2 \partial^\mu \chi \partial_\mu \chi$; then $X_2$ and one $\partial_\mu\sigma$ or 
$\partial_\mu\chi$ must be on internal lines, while the remaining $\partial^\mu\sigma$ or 
$\partial_\mu\chi$ leg is external. On the other hand, the constraint interaction vertex $\square X \sigma^2$ induces a contribution with two derivatives on the internal line connected to the field $X$. There are two possibilities: either the incoming propagator is $\Delta_{XX}$ or $\Delta_{X\sigma}$ and then the vertex is harmless, since (see Appendix~A) these propagators fall off as $p^{-4}$ and therefore the two derivatives from the vertex are compensated by two of the inverse powers of the propagator (the net effect being equivalent to the insertion of a derivative-free vertex); or the propagator is $\Delta_{X_2 X}$, in which case the second internal line must a $\sigma$-line or a $\chi$-line, so that in order to maximize the UV degree of divergence of the graph this line must be connected with an interaction vertex induced by the dimension-6 operator $X_2 (D^\mu \phi)^\dagger D_\mu \phi$, which, in turn, can only increase the UV degree of divergence by $1$, as explained above.

All in all, the insertion of one $\square X \sigma^2$ (resp.~one $\square X \chi^2$) vertex in association with $X_2 \partial^\mu \sigma \partial_\mu \sigma$ 
(resp.~$X_2 \partial^\mu \chi \partial_\mu \chi$)
is equivalent to the insertion of $\sigma \partial_\mu \sigma$ (resp.~$\chi \partial_\mu \chi$)  and a decrease of the UV degree of divergence of the $n$-loop vacuum topology by $1$. At this point it becomes clear how to obtain the power-counting bounds at order $n$ in the loop expansion: as in $D=4$ at one loop one gets $\delta_1=4$, we  can accommodate at most four insertions of $\sigma \partial_\mu \sigma$,
$\chi \partial_\mu \chi$. Hence the highest possibly UV divergent 1-PI ancestor amplitudes have dimension $12$, and come from $(\sigma \partial_\mu \sigma)^4$
or $(\chi \partial_\mu \chi)^4$. Similarly, at order $n$ the highest possibly UV divergent 1-PI ancestor amplitudes have dimension $3 \delta_n$, and come from $(\sigma \partial_\mu \sigma)^{\delta_n}$
or $(\chi \partial_\mu \chi)^{\delta_n}$ insertions.

Consider now the $T_1\neq0$ sector. This source couples to a Higgs or Goldstone particle through the vertices $T_1 (\partial^\mu \sigma \partial_\mu \sigma + \partial^\mu \chi \partial_\mu \chi)$; thus,  on every UV divergent ancestor amplitude which contains diagrams with $\sigma$ or $\chi$ internal lines, one can perform the insertion of an arbitrarily high number of $T_1$-external sources without affecting its UV behaviour at $T_1=0$. These insertions can however be resummed, the simplest example being provided, as we shall soon show, by the one-loop field redefinition amplitude $\G^{(1)}_{\chi^* \omega}$.

\section{\label{sec:sol.and.stab}Solution and stability of the functional identities}

The theory functional identities translate at the quantum level in the corresponding relations for the vertex functional $\G$. This result holds as a consequence of the absence of anomalies for the gauge group at hand and provided that quantization is carried out according to the local subtraction of counterterms as prescribed by the Bogolubov R-operation, consistently order by order in the loop expansion~\cite{Velo:1976gh}. Notice that since all propagators are of the Klein-Gordon type, the QAP~\cite{Lam:1972mb,Breitenlohner:1977hr,Zimmermann:1969jj,Zimmermann:1972te} holds, ensuring that the possible breaking of the ST identity is a local functional in the fields and external sources of the theory. One can then apply standard methods of Algebraic Renormalization~\cite{Piguet:1995er} in order to prove in a regularization-independent way that the quantum vertex functional $\Gamma$ does indeed obey the defining symmetries of the model. In the following we will obtain the most general solutions to the model's functional identities whose classical approximation is given in~\1eq{b.eq} through~\noeq{sti}. 

\subsection{$b$ and constraint equations}

Let $\G^{(n)}$ denotes the coefficient of order $n$ in the loop expansion of $\G$; then the $b$-equation~\noeq{b.eq} and the constraint antighost and ghost equations~\noeq{constr.antigh.eq} read at order $n \geq 1$
\begin{align}
	&\frac{\delta \G^{(n)}}{\delta b} = 0; & &\frac{\delta \G^{(n)}}{\delta \bar c} = 0;& \frac{\delta \G^{(n)}}{\delta c} = 0,
	\label{b.eq.loop}
\end{align}
stating that the only dependence of $\G$ on $b, \bar c$  and $c$ enters at the classical level. 

\subsection{Antighost equation}

The antighost equation~\noeq{antigh.eq} at order $n \geq 1$ is 
\begin{align}
	\frac{\delta \G^{(n)}}{\delta \bar \omega} = \frac{ev}{\xi} \frac{\delta \G^{(n)}}{\delta \chi^*},
	\label{ag.eq.loop}
\end{align}
that is, $\G^{(n)}$ depends on $\bar \omega$ only via the combination 
\begin{align}
	\widetilde{\chi}^* = \chi^* + \frac{ev}{\xi} \bar \omega.
	\label{chitilde}	
\end{align}

\subsection{$X_{1,2}$ equations}

The $X_1$- and $X_2$-equations~\noeq{X1.eq} and~\noeq{X2.eq} for $\G^{(n)}$ read
\begin{align}
	\frac{\delta \G^{(n)}}{\delta X_1} &= \frac{1}{v} (\square + m^2) \frac{\delta \G^{(n)}}{\delta \bar c^*};& 
	\frac{\delta \G^{(n)}}{\delta X_2} &= \frac{1}{v} (\square + m^2) \frac{\delta \G^{(n)}}{\delta \bar c^*} +
	\frac{g}{\Lambda} \frac{\delta \G^{(n)}}{\delta T_1},
	\label{X.eqs.loops}
\end{align}
thus implying that the whole dependence on $X_1$ and $X_2$ can only arise through the combinations 
\begin{align}
	&\tbarc = \bar c^* + \frac{1}{v} (\square + m^2) (X_1+ X_2);&  &{\cal T}_1 = T_1 + \frac{g}{\Lambda} X_2.&
	\label{X2.subst}
\end{align}
In particular, \1eq{X.eqs.loops} tells us that the 1-PI amplitudes involving at least one $X_1$ or $X_2$ external legs are uniquely fixed in terms of amplitudes involving neither $X_1$ or $X_2$, from which it follows the remarkable fact that in the $X$-theory the substitutions in \1eq{X2.subst} do not get renormalized. 

\subsection{ST identity: general cohomological considerations}

Finding the solution to the ST identity is more involved since this equation is bilinear in the vertex functional. Let us start by assuming that the ST identity has been fulfilled up to order $n-1$; then, at order $n$ in the loop expansion, the breaking term for the regularized vertex functional $\G_R$
\begin{align}
{\cal S}_0 (\G_R^{(n)})  + \sum_{k=1}^{n-1}  \int \!\mathrm{d}^4 x \, \Big [ \frac{\delta \G^{(n-k)}}{\delta \sigma^*} \frac{\delta \G^{(k)}}{\delta \sigma} + \frac{\delta \G^{(n-k)}}{\delta \chi^*} \frac{\delta \G^{(k)}}{\delta \chi} \Big ] \equiv {\cal B}_R^{(n)}
\label{n.st.id}
\end{align}
is a local functional of ghost number 1 in the sense of formal power series, as a consequence of the QAP. In the equation above ${\cal S}_0$ denotes the linearized ST operator 
\begin{align}
	{\cal S}_0  (Y) & = \int \!\mathrm{d}^4 x \, \Big [ \partial_\mu \omega \frac{\delta Y}{\delta A_\mu}  + \frac{\delta \G^{(0)}}{\delta \chi^*} \frac{\delta Y}{\delta \chi} + \frac{\delta \G^{(0)}}{\delta \sigma^*} \frac{\delta Y}{\delta \sigma}
+ b \frac{\delta Y}{\delta \bar \omega}  \nonumber \\
&  + \frac{\delta \G^{(0)}}{\delta \sigma} \frac{\delta Y}{\delta \sigma^*} + \frac{\delta \G^{(0)}}{\delta \chi} \frac{\delta Y}{\delta \chi^*} \Big ] \nonumber \\
& = s Y + \int \!\mathrm{d}^4 x \, \Big [ \frac{\delta \G^{(0)}}{\delta \sigma} \frac{\delta Y}{\delta \sigma^*} + \frac{\delta \G^{(0)}}{\delta \chi} \frac{\delta Y}{\delta \chi^*}  \Big ],
\end{align}
which acts as the BRST differential $s$ on the fields of the theory while mapping the antifields into the classical equations of motion of their corresponding fields. In particular, ${\cal S}_0$ is nilpotent as a consequence of the validity of the ST identity for $\G^{(0)}$, as can be checked by direct computation.

Now, the Wess-Zumino consistency condition~\cite{Wess:1971yu} (or equivalently the Jacobi identity for the BV bracket \cite{Gomis:1994he}) ensures that ${\cal B}_R^{(n)}$ is ${\cal S}_0$-invariant
\begin{align}
    {\cal S}_0({\cal B}_R^{(n)}) = 0.
\end{align}
Thus, one is faced with the problem of computing $H({\cal S}_0|\mathrm{d})$ that is the cohomology modulo $\mathrm{d}$ of the linearized ST operator ${\cal S}_0$ in the sector of ghost number 1 in the variables $A_\mu,\sigma,\chi,\sigma^*,\tilde\chi^*, \omega$ and the BRST-invariant sources $\tbarc$, ${\cal T}_1$. Notice that one can neglect the dependence on $\bar c$, $c$ (since these are free fields) as well as $b$, which only enters at tree level. This is consistent with the fact that the ${\cal S}_0$-transformation of the shifted $\tilde \chi^*$-antifield is $b$-independent:
\begin{align}
    {\cal S}_0(\widetilde \chi^*) = \left . \frac{\delta \G^{(0)}}{\delta \chi} \right |_{b=0} .
    \label{chitilde.eq}
\end{align}
The cohomology $H({\cal S}_0|\mathrm{d})$ at ghost number $1$ is known to be empty for the (non-anomalous) Abelian group~\cite{Barnich:2000zw}. This means that there must exists a functional $\Upsilon^{(n)}$ such that
\begin{align}
    {\cal S}_0(\Upsilon^{(n)}) = -{\cal B}^{(n)}_R,
\end{align}
and therefore the $n$-th order symmetric vertex functional is given by
\begin{align}
    \G^{(n)} = \G^{(n)}_R + \Upsilon^{(n)} + {\cal I}^{(n)}, 
\end{align}
with ${\cal I}^{(n)}$ a ${\cal S}_0$-invariant functional of ghost number zero fixing the finite $n$-th order counter-terms of the model. As a consequence of the nilpotency of ${\cal S}_0$, ${\cal I}^{(n)}$ will decompose into: 
\begin{align}
	{\cal I}^{(n)} = {\cal I}^{(n)}_\mathrm{gi} + {\cal S}_0(Y^{(n)})
\end{align}
where ${\cal I}^{(n)}_\mathrm{gi}$ is a gauge-invariant local formal power series in the field strength and their derivatives, the field $\phi$  and their covariant derivatives and the BRST-invariant external sources of the theory; while $Y^{(n)}$ is an antifield-dependent functional governing the (generalized) finite field redefinitions at order $n$ in the loop expansion. In particular in the sector linear in the antifields the functional $Y^{(n)}$ yields 
\begin{align}
	{\cal S}_0\Big ( \int \!\mathrm{d}^4 x \, \Big [ \sigma^* f_\sigma(\sigma,\chi,A_\mu) + \chi^* f_\chi(\sigma,\chi,A_\mu)  \Big ] \Big) &= 
	\int \!\mathrm{d}^4 x \, \Big [  f_\sigma(\sigma,\chi,A_\mu) \frac{\delta \G^{(0)}}{\delta \sigma} + f_\chi(\sigma,\chi,A_\mu)   \frac{\delta \G^{(0)}}{\delta \chi}\nonumber \\
	& - \sigma^* s  f_\sigma(\sigma,\chi,A_\mu) - \chi^* s f_\chi(\sigma,\chi,A_\mu)  \Big ], 
\end{align}
which will induce the field redefinitions $\sigma \rightarrow \sigma +  f_\sigma(\sigma,\chi,A_\mu)$, $\chi \rightarrow \chi + f_\chi(\sigma,\chi,A_\mu)$. The latter, as already remarked before, will not necessarily be linear in the quantized fields. 

Notice, finally, that ${\cal I}^{(n)}$ and $Y^{(n)}$ must be compatible with the bounds set by the $n$-th order power-counting previously derived.

\subsubsection{One-loop field redefinitions\label{1lfr}}

\begin{figure}
    \centering
    \includegraphics[scale=.75]{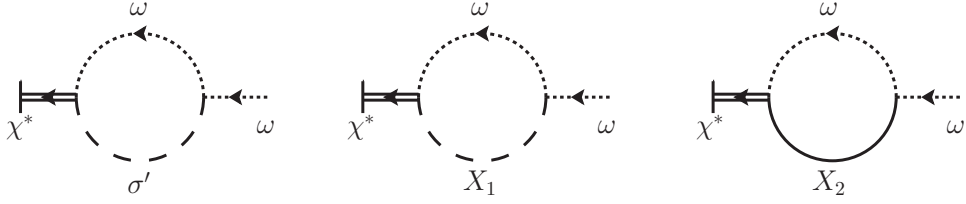}
    \caption{Diagrams contributing to the UV divergent 1-PI amplitude $\G^{(1)}_{\chi^* \omega}$.}
    \label{fig.2}
\end{figure} 

The field redefinitions compatible with the defining symmetries of the theory are very constrained by the ST identity, as the one-loop calculation we carry out in the following will show. We will use dimensional regularization around $D=4$. 

There is a unique UV divergent amplitude in the antifield sector at $T_1=0$, namely\footnote{In order to avoid notational clutter when dealing with 1-PI or connected amplitudes, we use subscripts to denote functional differentiation w.r.t. the arguments when setting afterwards to zero the fields and external sources. We also denote by a bar the UV divergent part of an amplitude (in dimensional regularization). For example, $\G^{(1)}_{\chi^* \omega} (x,y)\equiv \left . \frac{\delta^2 \G^{(1)}}{\delta \chi^*(x) \delta \omega(y)} \right |_{\Phi=\zeta=0}$, while ${\overline{\G}}^{(1)}_{\chi^* \omega}(x,y)$ is the pole part in the Laurent expansion around $D=4$ of the amplitude.} the function $\G^{(1)}_{\chi^* \omega}$ (the Feynman diagrams contributing to it are shown in \fig{fig.2}). A direct calculation yields
\begin{align}
	{\overline{\G}}^{(1)}_{\chi^* \omega}(x,y)  = \frac{e^2 M_A}{16 \pi^2} \frac{2}{4-D} \delta^{(4)}(x-y).
\end{align}

The $T_1$-dependent corrections to this Green's functions are obtained by repeated insertions of the vertex $\frac{T_1}{2} \partial_\mu \sigma \partial^\mu \sigma$ on the scalar propagators $\Delta_{\sigma'\sigma'}, \Delta_{X_1 X_1}, \Delta_{X_2 X_2}$ (see~\fig{fig.2} again). Since $\overline{\G}^{(1)}_{\chi^* \omega}$ is logarithmically divergent, one only needs to consider zero-momentum insertions of $T_1$. The final result is at $T_1 \neq 0$ 
\begin{align}
	\overline{\G}^{(1)}_{\chi^*\omega}[T_1](x,y) = \frac{1}{1+T_1(y)}  \frac{e^2 M_A}{16  \pi^2} \frac{2}{4-D} \delta^{(4)}(x-y).
\end{align}

On the other hand, the amplitudes $\G^{(1)}_{\chi^* \omega \sigma}$, $\G^{(1)}_{\sigma^* \omega \chi}$ are UV convergent. This is consistent with the parameterization of the UV divergences in the sector spanned by the antifields $\sigma^*, \chi^*$ in terms of the ${\cal S}_0$-exact invariant
\begin{align}
 \int \!\mathrm{d}^4 x \,   \frac{e M_A}{16 v \pi^2} \frac{2}{4-D} {\cal S}_0 \Big ( \frac{1}{1+T_1} ( \sigma^* \sigma + \chi^* \chi) \Big ) \supset - \int \!\mathrm{d}^4 x \, \frac{e^2 M_A}{16 \pi^2} \frac{2}{4-D}  \frac{1}{1+T_1}  \chi^* \omega \, .
\end{align} 
The above equation has some deep implications. First of all it states that in the $X$-theory the only allowed field redefintion at one loop level is linear in the scalar fields, with a prescribed dependence on the source $T_1$:
\begin{align}
	&\sigma \rightarrow \sigma + \frac{c^{(1)}}{1+T_1} \sigma;& &\chi \rightarrow \chi + \frac{c^{(1)}}{1+T_1} \chi;& &c^{(1)} = \frac{e M_A}{16 v \pi^2} \frac{2}{4-D}.
	\label{1loop.f.red}
\end{align}
Moreover in the sector spanned by the antifields the dependence on $T_1$ fixes in a unique way  the dependence on $X_2$ via the $X_2$-equation. Upon the mapping to the ordinary $\phi$-formalism this predicts the one loop field redefinitions in the target theory, as we will discuss in detail in Sect.~\ref{sec:mapping}. As we will show, such a field redefinition is no more linear in the target theory field variables, at variance with the power-counting renormalizable model as well as the non power-counting renormalizable theories where only derivative-independent scalar potentials are added~\cite{Binosi:2017ubk}.

One might also consider other possible dimension 6 operators with one $X_2$ field and two derivatives. The operator $\int \!\mathrm{d}^4 x \, X_2 \square (\phi^\dagger \phi - \frac{v^2}{2} )$ is controlled by the $X_2$-equation via the derivative with respect to $\bar c^*$. The remaining two operators
\begin{align}
	&\frac{g_2}{\Lambda} \int \!\mathrm{d}^4 x \, X_2 \partial^\mu (\phi^\dagger D_\mu \phi + (D_\mu \phi)^\dagger \phi);& &\frac{g_3}{\Lambda} \int \!\mathrm{d}^4 x \,  X_2 \phi^\dagger D^2 \phi,
\end{align}
can be safely introduced into the classical action
by coupling them to additional sources $T_i$, $i=2,3$.
The $X_2$-equation gets modified as follows (we set $g_1=g$):
\begin{align}
\frac{\delta \G^{(0)}}{\delta X_2} =  \frac{1}{v} (\square + m^2) \frac{\delta \G^{(0)}}{\delta \bar c^*} + \sum_{i=1}^3 \frac{g_i}{\Lambda} \frac{\delta \G^{(0)}}{\delta T_i} - (\square + m^2)X_1 - (\square + M^2) X_2 - v \bar c^* \, .
\label{X2.eq.mod}
\end{align}
The solution to the $X_2$-equation is trivially modified and its most general solution is recovered by making use of the replacements ${\cal T}_i = T_i + \frac{g_i}{\Lambda} X_2$, in addition to the $\bar c^*$-substitution in Eq.(\ref{X2.subst}).

The  insertion of the vertex $T_2 \partial^\mu (\sigma \partial_\mu \sigma)$ on the $\sigma',X_1,X_2$-lines of the amplitude $\G^{(1)}_{\chi^* \omega}$  does not contribute to the UV divergent coefficient of $\G^{(1)}_{\chi^* \omega}[T_i]$. This is because one derivative acts on the external source $T_2$ and thus the insertion of the $T_2$-vertex amounts to an increase by one of the UV degree  (due to the derivative acting on the internal $\sigma',X_1,X_2$-fields) and  a decrease by two due to the insertion of an additional scalar propagator, so that overall the UV degree decreases by one, thus leaving a UV convergent amplitude. The insertion of $\frac{T_3}{2} \sigma \square \sigma$ yields instead
\begin{align}
\overline{\G}^{(1)}[T_i]_{\chi^* \omega} \supset \frac{1}{1-T_3(y)} c^{(1)} \delta^{(4)}(x-y).
\end{align}

Let us finally compare this list of operators with the one given in Table~2 of Ref.~\cite{Grzadkowski:2010es}. The operator $Q_{\phi \square} \equiv (\phi^\dagger \phi) \square (\phi^\dagger \phi)$ trivially corresponds to 
$X_2 \square (\phi^\dagger \phi - \frac{v^2}{2})$, while
$Q_{\phi D} \equiv (\phi^\dagger D^\mu \phi)^\dagger (\phi^\dagger D^\mu \phi)$ in the Abelian case reduces to $\phi^\dagger \phi (D^\mu \phi)^\dagger D_\mu \phi$, which we decompose
as the sum 
\begin{align}
    \phi^\dagger \phi (D^\mu \phi)^\dagger D_\mu \phi & = 
\Big ( \phi^\dagger \phi - \frac{v^2}{2} \Big ) (D^\mu \phi)^\dagger D_\mu \phi  + \frac{v^2}{2} (D^\mu \phi)^\dagger D_\mu \phi \nonumber \\
& \Rightarrow X_2 (D^\mu \phi)^\dagger D_\mu \phi  + \frac{v^2}{2} (D^\mu \phi)^\dagger D_\mu \phi
\, .
\end{align}
Notice that this separation stays meaningful also at the quantum level, since the $X_1$-equation implies that the insertion of $\phi^\dagger \phi - \frac{v^2}{2} - v X_2$, i.e. the operator coupled to $\bar c^*$, is indeed the insertion of the null operator on physical amplitudes to all orders in the loop expansion.

\subsection{ST identity: bleaching}

Let us now show how to explicitly solve the iteratively imposed ST identity 
\begin{align}
{\cal S}_0 (\G^{(n)})=\Delta^{(n)}= - \sum_{k=1}^{n-1}  \int \!\mathrm{d}^4 x \, \Big [ \frac{\delta \G^{(n-k)}}{\delta \sigma^*} \frac{\delta \G^{(k)}}{\delta \sigma} + \frac{\delta \G^{(n-k)}}{\delta \chi^*} \frac{\delta \G^{(k)}}{\delta \chi} \Big ].
\label{n.st.id.imp}
\end{align}
Notice in particular that the results presented here will be valid for the full (non-local) vertex functional $\G^{(n)}$ and not limited to its local approximation (which controls the counterterms). 

Using the change of variable~\noeq{chitilde} allows us to set $b=0$ due to the validity of \1eq{chitilde.eq}. Moreover we can also set $X_{1,2}=0$: the dependence on these fields can be in fact recovered via \2eqs{X1.eq}{X2.eq} and the associated replacements~\noeq{X2.subst}, since $X_1$, $X_2$, $\bar c^*$, $T_1$ are ${\cal S}_0$-invariant.

Next we will now perform an invertible change of variables on the gauge field $A_\mu$ and the Higgs field $\sigma$ in order to obtain a new set of ${\cal S}_0$-invariant (and gauge-invariant) fields which we will name bleached fields. The method is an extension of the one originally devised for the Algebraic Renormalization of nonlinarly realized gauge theories~\cite{Bettinelli:2007tq,Bettinelli:2007cy,Bettinelli:2008ey,Bettinelli:2008qn} and amounts to a field-dependent finite gauge transformation. For this purpose we introduce the scalar
\begin{align}
\Omega = \frac{\phi}{\sqrt{\phi^\dagger \phi}} \equiv \Omega_0 + i \Omega_1 \, , \qquad \Omega_0 = \frac{\sigma+v}{\sqrt{(\sigma+v)^2+\chi^2}} \, , \quad  \Omega_1 = \frac{\chi}{\sqrt{(\sigma+v)^2+\chi^2}} \, .
\end{align}
The normalization is chosen in such a way that $\Omega^\dagger \Omega = 1$ (at $\chi=0$ one has $\Omega=1$); finally, $\Omega$ transforms as $\phi$ under the U(1) gauge symmetry.

The bleached fields are then defined as
\begin{align}
	\widetilde{A}_\mu &=  A_\mu + \frac{i}{e} \Omega^\dagger \partial_\mu \Omega=A_\mu - \frac{1}{e} \frac{1}{(\sigma+v)^2+\chi^2} \left[ (\sigma + v) \partial_\mu \chi - \chi \partial_\mu \sigma \right],\nonumber \\
	  \widetilde{\phi} &= \Omega^\dagger \phi=\sqrt{(\sigma + v)^2 + \chi^2} - v.
	\label{blch.vars}
\end{align}
Notice that the imaginary component of $\widetilde{\phi}$ identically vanishes; and, as anticipated, at $\chi=0$ one recovers the original fields: $\widetilde{A}_\mu = A_\mu$ and $\widetilde \sigma = \sigma$. 
In terms of these new fields the l.h.s. of the ST identity~\noeq{n.st.id.imp} reads
\begin{align}
 	{\cal S}_0(\G^{(n)}) =\int \!\mathrm{d}^4 x \, \Big [ e \omega (\sigma + v) \frac{\delta \G^{(n)}}{\delta \chi} + \frac{\delta \G^{(0)}}{\delta \sigma} \frac{\delta \G^{(n)}}{\delta \sigma^*} + \frac{\delta \G^{(0)}}{\delta \chi} \frac{\delta \G^{(n)}}{\delta \widetilde \chi^*} \Big ].
 \label{st.bl}
\end{align}
Observe now that 
\begin{align}
	\frac{\delta \G^{(0)}}{\delta \sigma} = v \bar c^* + \cdots,
\end{align}
so that it is advantageous to redefine $\bar c^*$ according to 
\begin{align}
   \bar c^*\to \tilde{\bar c}^* = \frac{\delta \G^{(0)}}{\delta \sigma}.
    \label{bc.redef}
\end{align}
Then ${\cal S}_0(\sigma^*) = \tilde{\bar c}^*$, ${\cal S}_0(\tilde{\bar c}^*)=0$: $(\sigma^*,\tilde{\bar c}^*)$ form a BRST doublet~\cite{Quadri:2002nh} (also called a contractible pair~\cite{Barnich:2000zw}), and it is known that doublets do not contribute to the cohomology of ${\cal S}_0$~\cite{Quadri:2002nh,Barnich:2000zw}. A final change of variables 
\begin{align}
    \omega \rightarrow \widetilde \omega = e \omega (\sigma + v),
    \label{ghost.chv}
\end{align} 
allows one to rewrite the above equation in its final form 
\begin{align}
	\int \!\mathrm{d}^4 x \,  \Big [ \widetilde\omega \frac{\delta \G^{(n)}}{\delta \chi} + \tilde{\bar c}^* \frac{\delta \G^{(n)}}{\delta \sigma^*} \Big ] = -
	\int \!\mathrm{d}^4 x \, \frac{\delta \G^{(0)}}{\delta \chi} \frac{\delta \G^{(n)}}{\delta \tilde\chi^*} - \Delta^{(n)}.
\label{st.bl.hom}
\end{align}

At this point we can write the left-hand side of the equation above in terms of an auxiliary nilpotent operator $\rho$ belonging to the class of differentials acting non-trivially only on BRST doublets, namely $(\chi,\tilde\omega)$ and $(\sigma^*,\tilde{\bar c}^*)$ for the case at hand: 
\begin{align}
	&\rho (\G^{(n)})\equiv\int \!\mathrm{d}^4 x \,  \Big [ \widetilde\omega \frac{\delta \G^{(n)}}{\delta \chi} + \tilde{\bar c}^* \frac{\delta \G^{(n)}}{\delta \sigma^*} \Big ]& \Longrightarrow & & \rho (\G^{(n)})=-
	\int \!\mathrm{d}^4 x \, \frac{\delta \G^{(0)}}{\delta \chi} \frac{\delta \G^{(n)}}{\delta \tilde\chi^*} - \Delta^{(n)},&
\end{align}
so that one can in principle get hold of $\Gamma^{(n)}$ provided he knows the inverse of the $\rho$ operator. On the other hand, $\rho$ can be easily inverted since it admits an associated homotopy operator ${\cal h}$~\cite{Wess:1971yu,Picariello:2001ri}, {\it i.e.}, an operator such that
\begin{align}
    \{ \rho, {\cal h} \} = \mathbb{I},
\end{align}
where $\{\cdot,\cdot\}$ indicates the usual anticommutator, while $\mathbb{I}$ is the identity on the space of functionals depending on the doublets. Indeed, let us define
\begin{align}
    {\cal h}(Y) = \int_0^1 \!\mathrm{d}t \! \int \!\mathrm{d}^4 x \, \Big [ \chi \lambda_t \frac{\delta Y}{\delta \widetilde \omega} + \sigma^* \lambda_t \frac{\delta Y}{\delta \tilde{\bar c}^*} \Big ], 
\end{align}
where the operator $\lambda_t$ acts on a functional $Y$ by multiplying the doublets it depends upon by $t$, leaving unaffected all the other variables $\xi$ on which $Y$ might depend:
\begin{align}
    \lambda_t Y(\chi,\tilde\omega,\sigma^*,\tilde{\bar c}^*;\xi) = 
    Y(t \chi,t \tilde\omega,t \sigma^*,t \tilde{\bar c}^*;\xi).
\end{align}
Then
it is straightforward to verify that
\begin{align}
    \{ \rho, {\cal h} \}Y = Y(\chi,\tilde\omega,\sigma^*,\tilde{\bar c}^*;\xi) - Y(0,0,0,0;\xi).
\end{align}
This identity holds
provided that the space of $Y$ functionals is star-shaped, {\it i.e.}, any of its elements rescaled by $t$ is still in $Y$; and this is certainly the case for the space of functionals depending on the fields and the external sources of the theory.

Moreover, as a consequence of the nilpotency of $\rho$, the r.h.s. of Eq.(\ref{st.bl.hom})
is $\rho$-invariant, and thus one finds
\begin{align}
    \{ \rho, {\cal h} \} \Big ( -
\int \!\mathrm{d}^4 x \, \frac{\delta \G^{(0)}}{\delta \chi} \frac{\delta \G^{(n)}}{\delta \tilde\chi^*} - \Delta^{(n)} \Big ) = \rho {\cal h} \Big ( -
\int \!\mathrm{d}^4 x \, \frac{\delta \G^{(0)}}{\delta \chi} \frac{\delta \G^{(n)}}{\delta \tilde\chi^*} - \Delta^{(n)} \Big ),
\end{align}
yielding the final representation for the $n$-th order vertex functional
\begin{align}
    \G^{(n)} = {\cal h} \Big ( - \int \!\mathrm{d}^4 x \frac{\delta \G^{(0)}}{\delta \chi} \frac{\delta \G^{(n)}}{\delta \tilde\chi^*} - \Delta^{(n)} \Big ) + \G^{(n)}_{\mathrm{ker}},
\label{nth.order.rep}
\end{align}
where $\G^{(n)}_{\mathrm{ker}}$ is a $\rho$-invariant functional built from the bleached variables $\tilde{A}_\mu,\tilde \sigma$, the invariant source $T_1$ and the pairs $(\chi, \tilde\omega)$,
$(\sigma^*,\tilde{\bar c}^*)$. Eq.(\ref{nth.order.rep})  provides a compact representation for the most general solution to the ST identity at the $n^\mathrm{th}$ loop order.
In particular in the ghost-independent sector $\G^{(n)}_{\mathrm{ker}}$ reduces to the gauge-invariant functional 
$$\G^{(n)}_{\mathrm{gi}} = \left . \G^{(n)}_{\mathrm{ker}} \right |_{\tilde{\omega}=0} \, .$$

Summarizing, \1eq{nth.order.rep} tells us that there are two possible contributions to amplitudes involving the antifields (and thus controlling the field redefinitions~\cite{Gomis:1994he,Gomis:1995jp}): those arising from lower order terms, namely from $\Delta^{(n)}$ in \1eq{nth.order.rep}; and those arising from the $n$-order amplitudes. In all cases, however, only amplitudes involving the external source $\widetilde{\chi}^*$ are relevant, since due to SSB the Higgs antifield $\sigma^*$ naturally pairs into a BRST doublet with the source $\tilde{\bar c}^*$. Then, the effect of field redefinitions on the $n$-th order amplitudes can be controlled by the homotopy operator ${\cal h}$; the remaining contributions in the sector
without ghosts are strictly gauge-invariant and represented by the functional $\G^{(n)}_{\mathrm{gi}}$, depending only on the bleached variables and the invariant external sources.

\subsubsection{Recovering the ST identities at one-loop.}

Let us see how the general solution of the ST identity~\noeq{nth.order.rep} works at the one-loop level. At this order in the loop expansion there is no contribution from $\Delta^{(n)}$ and one therefore has
\begin{align}
    \G^{(1)}= {\cal h} \Big ( - \int \!\mathrm{d}^4 x \frac{\delta \G^{(0)}}{\delta \chi} \frac{\delta \G^{(1)}}{\delta \tilde\chi^*} \Big ) + \G^{(1)}_\mathrm{ker}.
    \label{l1.st.sol}
\end{align}
In the following we are going to check that indeed~\1eq{l1.st.sol} gives the usual ST identities for the two point functions in the $\chi, A_\mu$ sector.

To evaluate the right-hand side of~\1eq{l1.st.sol} we first need to write down the equations of motion for the $\chi,\sigma$ fields in terms of the bleached variables (the $\sigma$-field equation is required for the redefinition of the $\tilde{\bar c}^*$ external source in~\1eq{bc.redef}). It is then advantageous to rewrite the $\chi,\sigma$ derivatives in terms of $\phi$ and its complex conjugate $\phi^\dagger$:
\begin{align}
    &\frac{\delta}{\delta \chi} = \frac{i}{\sqrt{2}} \Big ( \frac{\delta}{\delta \phi} - \frac{\delta}{\delta \phi^\dagger} \Big );&
    &\frac{\delta}{\delta \sigma} =
    \frac{1}{\sqrt{2}} \Big ( \frac{\delta}{\delta \phi} + \frac{\delta}{\delta \phi^\dagger} \Big ).
\end{align}
By acting with the above differential operators on $\G^{(0)}$ gauge covariant quantities arise; for instance let us consider the $\chi$ derivative of the covariant kinetic term, in which case one has
\begin{align}
    \frac{\delta}{\delta \chi(x)} \int \!\mathrm{d}^4 y \, 
    (D_\mu \phi)^\dagger D^\mu \phi = 
    \frac{i}{\sqrt{2}}
    \left[ - (D^2 \phi)^\dagger(x) + (D^2\phi)(x) \right].
    \label{sample.1}
\end{align}
According to the prescription for the homotopy evaluation, the right-hand side of the above equation needs to be rewritten in terms of bleached variables. Since this change of variables is a finite operatorial gauge transformation, this amounts to the multiplication by $\Omega$ of all quantities transforming in the fundamental representation of the gauge group and by $\Omega^\dagger$ for those transforming in the complex conjugate representation; indeed, in agreement with~\1eq{blch.vars}
\begin{align}
    &\phi = \Omega \widetilde{\phi};&
    &\phi^\dagger = \widetilde{\phi}^\dagger \Omega^\dagger;&
    &A_\mu = \widetilde{A}_\mu - \frac{i}{e} \Omega^\dagger \partial_\mu \Omega,
    \label{bl.ch1}
\end{align}
from which~\1eq{sample.1} becomes
\begin{align}
     \frac{\delta}{\delta \chi(x)} \int \!\mathrm{d}^4 y \, 
    (D_\mu \phi)^\dagger D^\mu \phi = 
    \frac{i}{\sqrt{2}}
    \left[ - \Omega^\dagger (\widetilde{D}^2 \widetilde{\phi})^\dagger(x) + \Omega (\widetilde{D}^2\widetilde \phi)(x) \right].
    \label{cov.kin.blch}
\end{align}
Finally one will need also to replace the ghost $\omega$ according 
to ~\1eq{ghost.chv}, that is
\begin{align}
    \omega = \frac{\tilde \omega}{e (\sigma + v)} =
    \frac{\tilde \omega}{e \sqrt{(\tilde{\sigma}+v)^2-\chi^2}}.
    \label{w.ch}
\end{align}

We are now in a position to derive the contributions to the two point $\chi\chi$ and $\chi A_\mu$ 1-PI amplitudes at one loop order.

We start from the homotopy term. By inspection one sees that the only term in the classical equation of motion of $\chi$, contributing to the $\chi A_\mu$ sector, appears in the covariant kinetic term via a contribution linear in the bleached gauge field
\begin{align}
	 \frac{i}{\sqrt{2}}
    \left[ - \Omega^\dagger (\widetilde{D}^2 \widetilde{\phi})^\dagger(x) + \Omega (\widetilde{D}^2\widetilde \phi)(x) \right]\supset  ev\, \partial^\mu \widetilde{A}_\mu,
\end{align}
so that the homotopy term yields
\begin{align}
	\int_0^1\!\mathrm{d}t\! \int \!\mathrm{d}^4 x \, \chi \lambda_t \frac{\delta}{\delta \tilde{\omega}} \Big [
    - \int \!\mathrm{d}^4 y \, \frac{\delta \G^{(0)}}{\delta \chi(y)} \frac{\delta \G^{(1)}}{\delta \tilde{\chi}^*(y)}\Big] & \sim 
    - \int_0^1\!\mathrm{d}t\!\int \! \mathrm{d}^4 x  \!\int \!\mathrm{d}^4 y\, ev \chi(x)  \partial^\mu{\widetilde{A}_\mu(y)} \lambda_t \times\nonumber \\
    &\times\frac{\delta^2 \G^{(1)}}{\delta \tilde \omega(x) \delta{\tilde \chi^*(y)}} \nonumber \\
& \sim - \int \! \mathrm{d}^4 x  \!\int \!\mathrm{d}^4 y\, \chi(x) \partial^\mu{\widetilde{A}_\mu(y)} \G^{(1)}_{\omega\chi^*}(x,y), 
\end{align}
where in the last line we have used the fact that, since  the pre-factor $\chi \partial^\mu{\widetilde A}_\mu$ is already of second order in the gauge and Goldstone fields, we can neglect the field dependence in~\1eq{w.ch} and use $\omega \sim\widetilde \omega/ev$. The last step is to convert back to the original variables by using the last of~\1eq{bl.ch1} to get the final expression
\begin{align}
\G^{(1)}_{\mathrm{hom}} \equiv -
   \int \! \mathrm{d}^4 x  \!\int \! \mathrm{d}^4 y\, \chi(x) \Big (\partial^\mu A_\mu - \frac{1}{ev} \square \chi \Big )(y) \G^{(1)}_{\omega\chi^*}(x,y).
   \label{hom}
\end{align}

We now move to the gauge-invariant contribution generated by $\G^{(1)}_\mathrm{gi}$. Only the two-point function of the bleached field can contribute to the two-point Goldstone-gauge sector, so that, exploiting the Bose symmetry of the two-point gauge function, we get
\begin{align}
    \G^{(1)}_\mathrm{gi} & \sim \int \! \mathrm{d}^4 x  \!\int \! \mathrm{d}^4 y\, 
    \frac{1}{2} \G^{(1)}_{A_\mu A_\nu}(x,y) \widetilde{A}_\mu(x)
    \widetilde{A}_\nu(y) 
    \nonumber \\
    & \sim
    \int \! \mathrm{d}^4 x  \!\int \! \mathrm{d}^4 y\,  
    \frac{1}{2} \G^{(1)}_{A_\mu A_\nu}(x,y) \Big[
        A_\mu(x)
        A_\nu(y)
        + \frac{1}{(ev)^2}\partial_\mu \chi(x)
        \partial_\nu \chi(y)
        -  \frac{2}{ev} \partial_\mu \chi(x) A_\nu (y)
    \Big].
    \label{gi}
\end{align}

\2eqs{hom}{gi} allows us to read off the $\chi\chi$ and $\chi A_\mu$ two-point functions by simply differentiating them with the relevant field combinations. One finds in the momentum space ($\partial_\mu\to ip_\mu$):
\begin{align}
    & \G^{(1)}_{\chi\chi} = \frac{1}{(ev)^2} \G^{(1)}_{A_\mu A_\nu} p^\mu p^\nu - \frac{2}{ev} p^2 \G^{(1)}_{\omega \chi^*};&
   & \G^{(1)}_{\chi A_\mu(p)} = - \frac{1}{ev} i p^\nu \G^{(1)}_{A_\nu A_\mu} + i p^\mu \G^{(1)}_{\omega \chi^*},
   \label{2pt}
\end{align}
where the first term in the r.h.s. arises from $\G^{(1)}_\mathrm{gi}$ and the second from $\G^{(1)}_\mathrm{hom}$. By noticing that $\G^{(0)}_{\omega  \chi^*} = ev$ and $\G^{(0)}_{\chi A_\mu(p)} = - i ev  p^\mu$, the second equation of~\noeq{2pt} can be recast into
\begin{align}
    i p^\nu \G^{(1)}_{A_\nu A_\mu(p)} + \G^{(0)}_{\omega  \chi^*} \G^{(1)}_{\chi A_\mu(p)} + \G^{(1)}_{\omega \chi^*} \G^{(0)}_{\chi A_\mu(p)}  = 0.
    \label{2pt.a.fin}
\end{align}
In addition, by multiplying the second equation in~\noeq{2pt} by $-i p^\mu$ and using the first expression in the same equation in order to substitute $p^\mu p^\nu \G^{(1)}_{A_\nu A_\mu(p)}$ we find (notice that $\G^{(0)}_{\chi\chi} = p^2$)
\begin{align}
    - i p^\mu \G^{(1)}_{\chi A_\mu(p)} +  \G^{(0)}_{\omega \chi^*} \G^{(1)}_{\chi\chi}  + 
	\G^{(1)}_{\omega \chi^*}\G^{(0)}_{\chi\chi}  = 0.
 	\label{2pt.b.fin}   
\end{align}
It is then strightforward to show \2eqs{2pt.a.fin}{2pt.b.fin} coincide with the ST identities obtained by differentiating~\1eq{sti}, taken at $n=1$,  with respect to the ghost $\omega$ and then either $A_\mu$ or $\chi$ respectively.

\section{Mapping to the $\phi$-representation}\label{sec:mapping}

In the $\phi$-representation the presence of the operator $\phi^\dagger \phi (D^\mu \phi)^\dagger D_\mu \phi$ generates vertices with three and four external $\sigma$-legs and two derivatives. They in turn give rise to one-loop amplitudes logarithmically divergent no matter how many external $\sigma$-legs are inserted. It is then very difficult to elucidate and classify the UV divergences of the model and separate the unphysical contributions, induced by generalized field redefinitions, from the genuine physical effects of such a higher dimensional operator.

On the contrary the task becomes rather straightforward if one makes use of the mapping of the 1-PI amplitudes in the $X$-formalism on those of the usual $\phi$-representation; for the purpose of deriving this mapping all one has to do is to just to go on-shell with the fields $X_{1,2}$. 

Let us show how the procedure works at the one-loop level, in which case it is enough to consider the tree-level equations of motion for these fields. More specifically, the $X_1$-equation of motion (\ref{X1EOM}) enforces the constraint $X_2 = \frac{1}{v} \Big ( \phi^\dagger \phi - \frac{v^2}{2} \Big )$. Once one takes into account this constraint, the $X_2$-equation of motion in turn yields 
\begin{align}
    (\square + m^2) (X_1 + X_2) = - (M^2 - m^2) X_2 + \frac{g}{\Lambda} (D^\mu \phi)^\dagger D_\mu \phi - v \bar c^*. 
    \label{X2.EOM}
\end{align}
By substituting the expressions for $X_{1,2}$ into the replacement rules~\noeq{X2.subst}
we obtain their final form (at zero external sources):
\begin{align}
 	& \tbarc \rightarrow - \frac{(M^2 - m^2)}{v^2} \Big ( \phi^\dagger \phi - \frac{v^2}{2} \Big ) + \frac{g}{v \Lambda} (D^\mu \phi)^\dagger D_\mu \phi;&
	& {\cal T}_1 \rightarrow  \frac{g}{v \Lambda} \Big ( \phi^\dagger \phi - \frac{v^2}{2} \Big ).
	\label{repl.fin}
\end{align}

Then we can start deriving some properties of the 1-PI amplitudes in the target theory by exploiting the (easier) renormalization of the model in the $X$-formalism.

\subsection{Field redefinitions}

We first discuss the field redefinitions (\ref{1loop.f.red}) under the mapping. 
One has to replace $T_1$ according to~\1eq{repl.fin}, thus obtaining
\begin{align}
    &\sigma \rightarrow \sigma + \frac{c^{(1)}}{1+ \frac{g}{\Lambda v} \Big ( \phi^\dagger \phi - \frac{v^2}{2} \Big )} \sigma; &   
    &\chi \rightarrow \chi + \frac{c^{(1)}}{1+ \frac{g}{\Lambda v} \Big ( \phi^\dagger \phi - \frac{v^2}{2} \Big )} \chi.
    \label{fr.target}
\end{align}

This is a quite remarkable result. It shows that there are indeed generalized field redefinitions in the target theory and that these field redefinitions are not even polynomial already at the one loop order. These redefinitions must therefore be properly taken into account if one wishes to renormalize the theory off-shell. Notice also that the rescaling factor of the complex Higgs field is gauge-invariant, as a consequence of the fact that the source 
${\cal T}_1$ is gauge-invariant.

\subsection{Two-point Higgs function}

Let us now complete our analysis of the two-point functions of the model by
considering the two-point Higgs function.
In order to obtain this amplitude via the mapping  we need to consider the
following 1-PI Green's functions in the $X$-theory.

\begin{itemize}
	\item The first amplitudes to be considered are the tapdoles $\G^{(1)}_{T_1}$ and $\G^{(1)}_{\bar c^*}$, yielding via the mapping (\ref{fr.target}) the following contributions
	\begin{subequations}
	\begin{align}
    	& \int \G^{(1)}_{T_1} T_1 \underset{\sigma\sigma \ \mathrm{term}}{\to} \int \frac{g}{2 \Lambda v} \G^{(1)}_{T_1} \sigma^2, \\
   	 & \int \G^{(1)}_{\bar c^*} \bar c^* \underset{\sigma\sigma \ \mathrm{term}}{\to} \int  \G^{(1)}_{\bar c^*}\Big ( -\frac{1}{2} \frac{M^2 - m^2}{v^2} \sigma^2 + \frac{1}{2} \frac{g}{\Lambda v} \partial^\mu \sigma \partial_\mu \sigma \Big ).    
	\end{align}
	\end{subequations}

	\item Next, we have amplitudes bilinear in the external sources:
	\begin{subequations}
		\begin{align}
			& \int \frac{1}{2} \G^{(1)}_{T_1(x) T_1(y)} T_1(x) T_1(y) \underset{\sigma\sigma \ \mathrm{term}}{\to} \int \frac{1}{2} \frac{g^2}{\Lambda^2} \G^{(1)}_{T_1(x) T_1(y)} \sigma(x) \sigma(y),\\
      		& \int \G^{(1)}_{T_1(x) \bar c^*(y)} T_1(x) \bar c^*(y) \underset{\sigma\sigma \ \mathrm{term}}{\to} -\int  \frac{g}{\Lambda v} (M^2-m^2) \G^{(1)}_{T_1(x) \bar c^*(y)} \sigma(x) \sigma(y), \\
      		& \int \frac{1}{2} \G^{(1)}_{\bar c^*(x) \bar c^*(y)} \bar c^*(x) \bar c^*(y) \underset{\sigma\sigma \ \mathrm{term}}{\to} \int \frac{1}{2} \frac{(M^2-m^2)^2}{v^2} \G^{(1)}_{\bar c^*(x) \bar c^*(y)} \sigma(x) \sigma(y).
		\end{align}
	\end{subequations}

	\item Finally, we need to consider the mixed $\sigma$-external sources amplitudes\footnote{Notice that one can safely replace $\sigma$ with $\sigma'$ since the transformation $\sigma'=\sigma - X_1 -X_2$ from the diagonal mass basis $(\sigma',X_1,X_2)$ to the symmetric basis $(\sigma,X_1,X_2)$ leaves the $\sigma,\sigma'$-amplitudes invariant.}
		\begin{subequations}
			\begin{align}
      			& \int \G^{(1)}_{T_1(x) \sigma'(y)} T_1(x) \sigma(y) \underset{\sigma\sigma \ \mathrm{term}}{\to} \int  \frac{g}{\Lambda}  \G^{(1)}_{T_1(x) \sigma'(y)} \sigma(x) \sigma(y), \\ 
      			& \int \G^{(1)}_{\bar c^*(x) \sigma'(y)} \bar c^*(x) \sigma(y) \underset{\sigma\sigma \ \mathrm{term}}{\to} -\int \frac{M^2-m^2}{v}  \G^{(1)}_{\bar c^*(x) \sigma'(y)} \sigma(x) \sigma(y),
      		\end{align}
		\end{subequations}
 		and the two point $\sigma$-amplitude
		\begin{align}
      		& \int \frac{1}{2} \G^{(1)}_{\sigma'(x) \sigma'(y)} \sigma(x) \sigma(y).
		\end{align}
\end{itemize}

Putting all the pieces together we obtain the following representation of the target two-point $\sigma$ function (we denote target amplitudes by a tilde)
\begin{align}
    \widetilde{\G}^{(1)}_{\sigma \sigma}(p^2) = & \G^{(1)}_{\sigma'\sigma'}  + 2 \frac{g}{\Lambda} \G^{(1)}_{T_1\sigma'} - 2 \frac{M^2-m^2}{v} \G^{(1)}_{\bar c^*\sigma'} \nonumber \\
    & + \frac{g}{\Lambda v} \G^{(1)}_{T_1} 
    + \Big ( - \frac{M^2-m^2}{v^2} + \frac{g}{\Lambda v} p^2 \Big ) \G^{(1)}_{\bar c^*}
    \nonumber \\
    & + \frac{g^2}{\Lambda^2}  \G^{(1)}_{T_1T_1}
    - 2 \frac{g}{\Lambda}\frac{M^2-m^2}{v} \G^{(1)}_{T_1 \bar c^*} + 
    \frac{(M^2-m^2)^2}{v^2} \G^{(1)}_{\bar c^* \bar c^*}.
    \label{2pt.sigma}
\end{align}
Notice that since~$m^2$ is an unphysical parameter~\cite{Binosi:2017ubk} the right-hand side cannot depend on it, as can be explicitly verified\footnote{This provides a very strong check of all the computations in the $X$-formalism, due to the generally large number of diagrams involved in them and the ubiquitous presence of $m^2$}. 

\1eq{2pt.sigma} provides a very useful decomposition of the two-point amplitude which cannot be obtained in a straightforward manner by a direct computation in the $\phi$-formalism. Terms in the first line are affected by the field redefinition via the $\sigma'$-dependence. On the other hand, terms in the second and third lines are unaffected by the field redefinitions being controlled by 1-PI amplitudes of external sources. In particular, the UV divergent term of order $p^4$ in $\widetilde{\G}^{(1)}_{\sigma \sigma}$ arises from the gauge-invariant amplitude $\G^{(1)}_{T_1 T_1}$, yielding
\begin{align}
    \frac{\partial}{\partial (p^2)^2} \overline{\widetilde{\G}}^{(1)}_{\sigma \sigma} 
    = \frac{g^2}{32\pi^2 \Lambda^2} \frac{1}{4-D}.
    \label{2pt.sigma.highest}
\end{align}
This term cannot be reabsorbed by a field redefinition $\sigma \rightarrow \sigma + \frac{1}{2} \frac{g^2}{32 \pi^2 \Lambda^2}\square \sigma$ since the allowed field redefinitions~\noeq{fr.target} do not contain derivatives. This means that~\1eq{2pt.sigma.highest} is a genuine contribution to a new physical operator not present in the classical action; since such contribution arises from the two-point function $\G^{(1)}_{T_1T_1}$, it is immediate to realize that the new operator is
\begin{align}
    \int \frac{1}{2} \overline{\G}^{(1)}_{T_1 T_1} T_1^2 \rightarrow 
    \int \!\mathrm{d}^4 x \, \frac{g^2}{v^2\Lambda^2} \frac{1}{64 \pi^2 (4-D)} \Big ( \phi^\dagger \phi - \frac{v^2}{2} \Big ) \square^2 \Big ( \phi^\dagger \phi - \frac{v^2}{2} \Big ).
\end{align}
%

\subsection{Renormalization of higher dimensional operators}

In the standard approach to EFTs, the perturbative series is organized in a separate expansion in the number of loops and (inverse) powers of the cutoff. More specifically, in the one-loop case, one treats multiple insertions of higher dimensional operators into one-loop diagrams as emerging at different orders in the cutoff expansion introducing at the same time the finite number of local counterterms that are required to cancel the divergences emerging up to that order. This would be straightforward if the only operators appearing in the parametrization of the corresponding UV divergences were gauge invariant. However, generalized field redefinitions associated to the cohomologically trivial invariants (and controlled by canonical transformations~\cite{Gomis:1994he,Gomis:1995jp}) appear. In the presence of SSB, these field redefinitions connect operators of different dimensions within the same order in the cutoff expansion\footnote{This can be explicitly seen in~\1eq{fr.target}, whose denominator contains the combination $sigma^2+\chi^2$ as well as the term $v\sigma$ both at first order in the inverse cutoff.}. Only after their presence has been properly accounted for, can one proceed to parametrize UV divergences by local counterterms associated to gauge invariant operators arranged in inverse powers of the cutoff. The number of these operators is infinite; however, at a given order in the cutoff expansion only a finite number of them will contribute, and one can in principle proceed at fixing their finite part by matching with the UV complete theory (in a given scheme). At this point, in the appropriate low energy regime, physical predictions only depend on those finite parts and not on the field redefinitions.     
 
 The procedure just described and, in particular, the last point, holds true iff the subtraction of the UV divergences has been done order by order in the loop expansion preserving the theory's defining functional identities such as the ST identity. This implies the correct identification of the cohomologically trivial invariants, otherwise one runs the risk of introducing spurious dependences on the latter in physical predictions, as well as, starting at two-loop order, the breaking of locality of the UV divergences. This will in turn uphold the standard assumption in the EFTs, that is the cancellation of higher dimensional UV divergences by higher dimensional local counterterms. And this is precisely what employing the formalism so far described avoids in a systematic fashion.      
 
In principle, the decomposition in~\1eq{l1.st.sol} is all one needs: the homotopy term can be computed explicitly and one can rather easily extract the gauge-invariant functional $\G^{(1)}_\mathrm{gi}$ containing all the relevant information required for the systematic evaluation of the one-loop $\beta$ functions of the theory. However, we face here a technical problem, appearing specifically in spontaneously broken gauge theories. In fact, the decomposition in~\1eq{l1.st.sol} gives a functional $\G^{(1)}_\mathrm{gi}$ that can be projected on monomials in the bleached fields and their ordinary derivatives. Then, when considering its local approximation (the relevant one for the evaluation of UV divergences and $\beta$ functions) this is achieved through a change of variables from the bleached coordinates to the basis of gauge invariant polynomials in the field strength and its derivatives and the gauge covariant field $\phi$ and its symmetrized covariant derivatives~\cite{Barnich:2000zw}.

This boils down to the solution of a linear system associated with a change of basis on a space of local functionals, bounded by the power counting (operators up to dimension 12 in the one-loop case at hand). While the solution is guaranteed to exist~\cite{Barnich:2000zw}, in practice one immediately faces the complication of a non-decoupling set of equations, due to the fact that the field $\phi$ exhibits a v.e.v., so that one has to solve the full tower of equations from the top. For instance, the dimension 12 operator  $(\phi^\dagger D^\mu \phi ~\phi^\dagger D_\mu \phi)^\dagger \phi^\dagger D^\nu \phi ~\phi^\dagger D_\nu \phi$ contributes to the four point $\sigma$-amplitude once one replaces each undifferentiated $\phi$ with its vev $v$:
\begin{align}
    (\phi^\dagger D^\mu \phi ~\phi^\dagger D_\mu \phi)^\dagger \phi^\dagger D^\nu \phi ~ \phi^\dagger D_\nu \phi \supset \frac{v^4}{16} \partial^\mu \sigma  \partial_\mu \sigma \partial^\nu \sigma  \partial_\nu \sigma.
\end{align}
Thus, if one insists in this way s/he cannot avoid the evaluation of all the multileg amplitudes required to solve the complete hierarchy of equations. 

There are, however, cases where the truncation of the linear system does occur, in a very subtle way. Consider, {\it e.g.}, the operator ${\cal O}(x) = F_{\mu\nu}^2 \Big ( \phi^\dagger \phi - \frac{v^2}{2}\Big)$: there cannot be contributions from field redefinitions to this operator, since the $\sigma$-field redefinition is derivative independent, see~\1eq{1loop.f.red}; and  no term in the classical action can generate this operator at $X_2=0$. Then, in order to fix the coefficient of ${\cal O}(x)$ it is enough to study the three point function ${\widetilde\G}^{(1)}_{A_\mu A_\nu \sigma}$. To prove this, let us briefly recall the contractible pairs change of variables leading to the familiar result that the gauge-invariant operators can be expressed as polynomials in the field strength and its derivatives and the field $\phi$ and covariant derivatives thereof~\cite{Barnich:2000zw}.

In the space of local functionals the appropriate coordinates are given by fields and their derivatives\footnote{We do not consider here the antifields, as done in Ref.\cite{Barnich:2000zw}, because for the particular operator at hand there are no contributions from field redefinitions.}, which are considered as independent coordinates (these are the coordinates in the so-called {\it jet space}). In cohomology computations it is convenient to introduce contractible pairs or BRST doublets, {\it i.e.}, pairs of variables $u,v$ such that $su=v$, $sv=0$; such pairs, in fact, do not contribute to the cohomology of the BRST differential.  The gauge field $A_\mu$ and its symmetrized derivatives form contractible pairs with the derivatives of the ghost; on the other hand, the field strength is, in the Abelian case, automatically BRST invariant. Hence, we can use the following variables
\begin{align}
     &\partial_{(\nu_1 \dots \nu_\ell}A_{\mu)};& &\partial_{(\nu_1 \dots \nu_\ell}\partial_{\mu)} \omega;&
     &\omega;&
     &\partial_{(\nu_1 \dots} \partial_{\nu_{\ell-1}} F_{\nu_\ell)\mu};& \quad D_{(\nu_1 \dots} D_{\nu_\ell)} \phi
\end{align}
where $(\dots)$ denote complete symmetrization, namely
\begin{align}
    D_{(\mu_1 \dots \mu_k)} = \frac{1}{k!} \sum_{\sigma \in S^k} D_{\mu_{\sigma(1)} \dots \mu_{\sigma(k)}},
\end{align}
with the sum is over the group $S^k$ of permutations of order $k$. Notice that the derivatives of the gauge field are recursively replaced by the contractible pairs by using the fact that 
\begin{align}
    \partial_{\nu_1 \dots \nu_\ell} A_\mu = \partial_{(\nu_1 \dots \nu_\ell} A_{\mu)}+\frac{\ell}{\ell+1}\partial_{(\nu_1 \dots \nu_{\ell-1}} F_{\nu_\ell) \mu},
    \label{cp.gauge}
\end{align}
which shows that the change of variables is invertible.

The procedure in order to evaluate the one-loop coefficient of the operator ${\cal O}(x)$ is then the following: evaluate the UV divergent part of ${\widetilde\G}^{(1)}_{A_\mu A_\nu \sigma}$ (which we denote by ${\widetilde{\overline{\G}}}^{(1)}_{A_\mu A_\nu \sigma})$. As explained above, no contributions from field redefinitions arise. Then use integration by parts in order to ensure that the $\sigma$ field is left undifferentiated in ${\widetilde{\overline{\G}}}^{(1)}_{A_\mu A_\nu \sigma}$. Indeed, monomials with undifferentiated gauge fields can be safely neglected since they form a contractible pair with the derivative of the ghost and thus they cannot affect the cohomology of ${\cal S}_0$ (they will cancel against contributions from other invariants containing the covariant derivatives of the $\phi$ field that do not concern us here; incidentally, this is the reason why one does not have to solve the full tower of equations in order to match the invariant expansions).

In order to recover these contributions by the by now familiar mapping technique, we need to evaluate three amplitudes in the $X$-formalism, namely $\G^{(1)}_{A_\mu A_\nu \sigma'}$, $\G^{(1)}_{A_\mu A_\nu T_1}$ and $\G^{(1)}_{A_\mu A_\nu \bar c^*}$. Then we reconstruct ${\widetilde\G}^{(1)}_{A_\mu A_\nu \sigma}$ according to\footnote{Again, the right-hand side of~\1eq{AAsigma} must be $m^2$ independent, which provides a strong check of the correctness of the computation.}
\begin{align}
    {\widetilde\G}^{(1)}_{A_\mu A_\nu \sigma} = \G^{(1)}_{A_\mu A_\nu \sigma'} - \frac{(M^2-m^2)}{v} \G^{(1)}_{A_\mu A_\nu \bar c^*} + \frac{g}{\Lambda} \G^{(1)}_{A_\mu A_\nu T_1}.
    \label{AAsigma}
\end{align}
A direct computation yields, after integration by parts in order to leave $\sigma$ derivative free, the following list of UV divergent terms:
\begin{align}
    \int \!\mathrm{d}^4 x\, \frac{1}{2} {\overline{\widetilde{\G}}}^{(1)}_{A_\mu A_\nu \sigma} A^\mu A^\nu \sigma &= \int \!\mathrm{d}^4 x \Big [ 
    \frac{r_1}{2} \partial^\mu A^\nu \partial_\mu A_\nu 
    + 
    \frac{r_2}{2} A^2  + \frac{r_3}{2} A_\mu \partial^\mu \partial^\nu A_\nu  + \nonumber \\
    & + \frac{r_4}{2} (\partial A)^2  + \frac{r_5}{2} A_\mu \square A^\mu \Big ] \sigma.
\end{align}
where the coefficients $r_i$ are collected in Appendix~\ref{app:F2sigma}.

Now, the only relevant monomial in order to obtain the coefficient
of ${\cal O}$ is the one associated to $r_1$, since, using~\1eq{cp.gauge} we can write
\begin{align}
	\partial^\mu A^\nu \partial_\mu A_\nu \sigma = 
	\partial^{(\mu} A^{\nu)}\partial_{(\mu} A_{\nu)} \sigma + 
	\frac{1}{4} F^{\mu\nu} F_{\mu\nu} \sigma \sim
	\frac{1}{4v} F^{\mu\nu} F_{\mu\nu} \Big ( \phi^\dagger \phi - \frac{v^2}{2} \Big ) + \cdots,
\end{align}
where the dots stand for additional operators that will not affect the operator under scrutiny. Thus, using the result~\noeq{r1},  we obtain that at one loop level the operator ${\cal O}(x)$ appears with the coefficient 
\begin{align}
	c^{(1)}_{\cal O} = \frac{r_1}{8v} = -\frac{1}{32 \pi^2} \frac{g^2 M_A^2}{\Lambda^2 v^2 }\frac{1}{4-D}.
\end{align}

In order to evaluate the $\beta$-function of this coupling (as well as of the other operators arising order by order in the loop expansion) the knowledge of the coefficients of all the invariant operators entering in $\overline{\Gamma}^{(1)}_{ \mathrm{gi}}$ is needed. Equivalently, one needs to know the running of the masses, of the vev and of the couplings (those already present at tree level and those radiatively generated). Once the  functional $\overline{\Gamma}^{(1)}_{ \mathrm{gi}}$  is known the problem boils down to carry out the appropriate change of variables to contractible pairs. However, as explained above, as a consequence of SSB this amount to solve a fairly complicated linear system involving multileg 1-PI amplitudes.
We will extensively report on the solution of such a linear system elsewhere.

As a final remark, we notice that in the EFT approach the required accuracy in  powers of the inverse cutoff for a given physical process constrains both the set of higher dimensional operators to be included in the classical action as well as the set of loop corrections to be evaluated. The formalism presented in this paper allows for a systematic and consistent treatment of the UV divergences subtractions for these processes: for each set of corrections one identifies the relevant loop order and carries out the renormalization of the theory, as described above, up to that order, by filtering the $n$-th loop coefficients of the invariants  according to the required inverse powers of the cutoff.

\section{Non-Abelian gauge theories}\label{sec:nonabelian}

An important question is whether the results obtained in this paper lend themselves to generalization in the context of non-Abelian gauge theories, {\it e.g.}, for the electroweak gauge group.

There are three key ingredients in the construction we have presented which can be summarized as follows: 
\begin{enumerate}
	\item The use of the gauge invariant combination $\phi^\dagger \phi - \frac{v^2}{2}$ as the dynamical variable describing the physical scalar mode, parameterized by the field $X_2$;
	\item The order-by-order solution of the ST identity via the bleaching technique, combined with the homotopy;
	\item The power-counting in the loop expansion.
\end{enumerate}
Let us examine each of them:
\begin{enumerate}
	\item The $X$-formalism implementing the description of SSB by the gauge-invariant dynamical variable $X_2$ can be straightforwardly generalized to the non-Abelian case; in fact both the $X_{1,2}$-equations and the constraint BRST symmetry hold for an arbitrary gauge group;
	\item The bleaching procedure can also be directly extended to a non-Abelian gauge group; in fact, the operatorial field redefinitions in~\1eq{blch.vars} have been written in a way that holds for an arbitary gauge group. Also the $\rho$ operator trivially generalizes to the non-Abelian setting (with the obvious sum over the $m$ pairs of Goldstone-ghost fields, $m$ being the dimension of the Lie algebra; {\it e. g.}, for $\rm SU(2)$ $m=3$, for $\rm SU(3)$ $m=8$). The pair $(\sigma^*,\tilde{\bar c}^*)$ does not change for a spontaneously broken non-Abelian gauge theory, so also the homotopy operator can be directly generalized;
	\item Finally, the power-counting too remains valid. In fact, the proof provided in Sect.~\ref{sec:pc} is based on the dimensions of the interaction vertices and does depend neither on the details of the field representations nor on the Abelian character of the gauge group.
\end{enumerate}

Of course the explicit off-shell renormalization of higher dimensional operators in a non-Abelian gauge theory requires some computational effort; however, the approach proposed here seems to be promising to tackle the off-shell renormalization problem of these theories too.

\section{Conclusions}\label{sec:conclusions}

In this paper we have addressed some aspects of the off-shell renormalization of an Abelian spontaneously broken gauge theory supplemented by dimension 6 derivative-dependent operators.

In the ordinary formalism (which we name $\phi$-representation) the classification of UV divergences for this model is very complicated already at one loop, due to the presence of an infinite set of divergent amplitudes with an arbitrary number of external $\sigma$-legs; and one cannot easily disentangle contributions from generalized field redefinitions from genuine physical effects arising from the renormalization of higher dimensional operators.

The use of the gauge invariant combination $\phi^\dagger \phi - \frac{v^2}{2}$ as the new dynamical variable describing the physical scalar mode ($X$-theory formalism) brings several advantages. First of all, a power-counting can be established for ancestor amplitudes ({\it i.e.}, those amplitudes which are not fixed by the functional identities of the model). In addition, amplitudes in the target theory get conveniently decomposed by the mapping from the $X$-theory to the $\phi$-representation according to the Green's functions of operators in the $X$-theory with definite properties under gauge transformations and generalized field redefinitions. In particular, one can explicitly obtain the one-loop field redefinition for the scalar $\phi$, that turns out to be not even polynomial; nevertheless its closed expression is easily obtained by exploiting the renormalization of the $X$-theory.

Next, we have proven that for spontaneously broken theories the ST identity can be explicitly solved by combining the bleaching of gauge and matter fields in the linear representation of the gauge group with homotopy techniques. This is a rather remarkable result in itself: Several attempts exist in the literature trying to directly impose the ST identity on the vertex functional ({\it e.g.}, with the aim of recursively fixing the finite counterterms required to restore the ST identity broken by intermediate regularization)~\cite{Ferrari:1998jy,Quadri:2003ui,Quadri:2003pq,Grassi:1999tp,Grassi:2000kp,Grassi:2001zz}. These attempts however focus only on the local approximation to the vertex functional (the relevant one when it comes to 
determine the finite counterterms restoring the possibly broken regularized  ST identity and to classify the UV divergences of the theory).
To the best of our knowledge, no general solution to all orders for the ST identity of spontaneously broken theories has been derived so far. 
For SMEFTs the main advantage of this solution is the constructive decomposition of the gauge-invariant part $\G^{(n)}_{\mathrm{gi}}$ from those unphysical terms removed by generalized field redefinitions, given by~\1eq{nth.order.rep}.

Finally we have identified the change of variables on the local approximation to $\G^{(n)}_\mathrm{gi}$ (relevant for the counterterms of the theory) from the Lorentz-invariant monomials in the bleached fields, the external sources and their ordinary derivatives thereof, to gauge-invariant polynomials in the field strengths, the gauge-covariant matter fields and their covariant derivatives; this is required in order to fully renormalize  the model. Albeit still a highly non-trivial task, what has been presented here paves the way for the systematic study of the off-shell renormalization of gauge-invariant operators; we will report on this subject in a forthcoming publication.

As a final comment, we would like to point out that it may happen that the $X$-theory formalism comes with some additional symmetries characterizing the special way it describes a SSB gauge theory. If that happens, it is likely to provide a deeper understanding of the $X$-theory approach to SSB.

\section*{Acknowledgments}

One of us (A.Q.) would like to thank ECT* and Fondazione Bruno Kessler for the warm hospitality. Useful discussions with G.Barnich are also acknowledged. Diagrams have been drawn using JaxoDraw~\cite{Binosi:2003yf,Binosi:2008ig}.

\appendix

\section{A toy scalar model}\label{app:toy}

That at a fixed loop order multiple insertions higher dimensional derivative operators can maximally violate power counting, thus leading to an infinite set of divergences, is a generic feature of EFTs. In this appendix we present a specific toy example constructed from a scalar theory possessing a trilinear and a quadrilinear non-renormalizable vertices, and discuss the advantages of formulating this model within the $X$-formalism.

Consider a massless ($M^2=0$) scalar theory supplemented by a non-renormalizable derivative interaction in four dimensions
\begin{eqnarray}
S = \int\! \mathrm{d}^4x \, \left[\frac{1}{2} \partial^\mu \sigma\partial_\mu \sigma 
+ \frac{g}{\Lambda^2} 
\Big ( \frac{1}{2} \sigma^2 + v \sigma \Big ) \partial^\mu \sigma \partial_\mu \sigma
\right].
\end{eqnarray}
In the ordinary formalism both the trilinear and the quadrilinear vertices give rise already at one loop order to an infinite number of UV divergent $\sigma$-amplitudes, due to the derivative nature of the interactions. Since external legs could be either $\sigma$ or $\partial\sigma$, and this influences the degree of divergence of the corresponding diagram, resummation of repeated insertion of, {\it e.g.}, the quadrlinear vertex on a UV divergent amplitude occurs only in particular subsets of diagrams: the ones where both $\partial\sigma$ are internal lines. The same happens for the trilinear vertices; in addition, in a SSB theory the two resummations are not independent, since the trilinear and quadrilinear couplings are related via the vev $v$. Thus, in order to identify the number of independent UV coefficients this requires to take into account such relations and deal effectively with the diagrams' combinatorial factors.

The $X$-formalism efficiently accomplishes both requirements. The vertex functional becomes (we omit the $m^2$ term, as it is immaterial for the ensuing discussion)
\begin{align}
    \G^{(0)} &= \int\! \mathrm{d}^4x \, \left[ 
    \frac{1}{2} \partial^\mu \sigma\partial_\mu \sigma  + 
    \frac{gv}{\Lambda^2} X_2 \partial^\mu \sigma \partial_\mu \sigma
     +\frac{1}{v} (X_1+X_2) \square \Big (\frac{1}{2} \sigma^2 + v \sigma - v X_2 \Big )\right. \nonumber \\
    &\left. + \bar c^* \Big ( \frac{1}{2} \sigma^2 + v \sigma - v X_2 \Big ) + T \partial^\mu \sigma \partial_\mu \sigma
    \right],
\end{align}
giving rise to propagators with the following UV behaviour (see also Appendix~\ref{app.prop})
\begin{align}
    &\Delta_{\sigma\sigma} \sim p^{-2};& 
    &\Delta_{\sigma X} \sim p^{-4};&
    &\Delta_{X X} \sim p^{-4},
\end{align}
where we set, as usual, $X=X_1+X_2$.

The defining functional $X$-equations of the model are then (see Sect.~\ref{sec:fids})
\begin{align}
    \frac{\delta \G^{(0)}}{\delta X_1} = 
    \frac{1}{v} \square \frac{\delta \G^{(0)}}{\delta \bar c^*} \, , \qquad
    \frac{\delta \G^{(0)}}{\delta X_2} =
    \frac{gv}{\Lambda^2} \frac{\delta \G^{(0)}}{\delta T} - \square (X_1 + X_2) - v \bar c^*,
\end{align}
while, going on-shell with $X_{1,2}$, gives rise to the one-loop mapping rules (see Sect.~\ref{sec:mapping})
\begin{align}
    &\bar c^* \rightarrow 
    \frac{g}{\Lambda^2} \partial^\mu \sigma \partial_\mu \sigma;&
    &T \rightarrow \frac{g}{\Lambda^2}
    \Big ( \frac{1}{2} \sigma^2 + v\sigma \Big ) .
    \label{toymap}
\end{align}

\begin{figure}
	\includegraphics[scale=0.75]{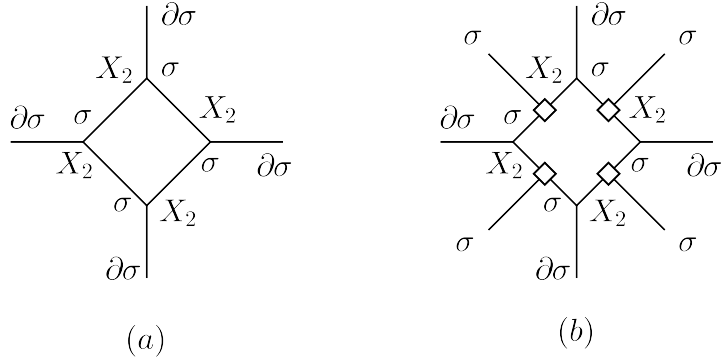}
	\caption{\label{fig:aapa} The most UV divergent one-loop $\sigma$-amplitudes in the $X$-theory formulation of the scalar toy-model.}
\end{figure}

The $X$-theory formulation then, displays the following UV properties:
\begin{enumerate}
    \item There is a manifest power-counting for the $\bar c^*$ and $\sigma$ amplitudes at $T=0$ external source. At one loop the UV divergent amplitudes involving $\bar c^*$ are limited to
    \begin{align}
        &\G^{(1)}_{\bar c^*};& 
        &\G^{(1)}_{\bar c^* \sigma};& 
        &\G^{(1)}_{\bar c^* \sigma \sigma};& 
        &\G^{(1)}_{\bar c^*\bar c^*}.
    \end{align}
    The reason why this happens is that: the interaction is now trilinear only; the $X_2$ line must be an internal line, as diagrams with external $X_2$ lines are obtained from the $X_2$-equation; and, consequently, $\partial\sigma$ is an external line. For one-loop divergent $\sigma$-amplitudes, the maximum number of external $\sigma$-legs is eight; additionally, the divergence must be of the form $(\sigma \partial^\mu \sigma)^4$, which is obtained from the box amplitude in Fig.~\ref{fig:aapa} ({\it a}) after insertion of the constraint vertex $\sim \square X \sigma^2$ [Fig.~\ref{fig:aapa} ({\it b})].

	\item For logarithmically UV divergent amplitudes the resummation of $T$-insertions is straighforward: it amounts to multiply the corresponding amplitude at $T=0$ by a pre-factor $1/(1+T)^n$, where $n$ is the number of $\sigma\sigma$ and $\sigma X$ propagators in the diagram; insertions on $XX$ propagators will decrease the UV degree of the amplitude by 2.
	
	\item The number of insertions of $T$ that one needs to consider depends on the dimension of the operators in the target theory one is interested in. In fact, under the mapping~\noeq{toymap} $T$ generates the operator $ \sigma^2/2 +v\sigma$; thus, if one is  interested, {\it e.g.}, in operators up to dimension $6$ in the target theory, one has to consider amplitudes with at most $3$ $T$-insertions on the power-counting-respecting amplitudes at $T=0$.
\end{enumerate}

Thus, the $X$-formalism streamlines the derivation of the independent UV coefficients and ensures that only the correct SSB field combination appears in the parametrization of UV divergences. The approach is even more advantageous when dealing with the intricacies of effective gauge theories, in particular in classifying the correct generalized field redefinitions, as explained in~Sect.~\ref{1lfr}.

\section{Propagators}\label{app.prop}

The diagonalization of the quadratic part of the classical action in the sector spanned by $\sigma,X_1,X_2$ is achieved via the field redefinition $\sigma=\sigma'+X_1+X_2$. In this new variables the propagators are
\begin{align}
    &\Delta_{\sigma'\sigma'} = \frac{i}{p^2-m^2}; &
    &\Delta_{X_1 X_1} = -\frac{i}{p^2-m^2};&
    &\Delta_{X_2X_2} = \frac{i}{p^2-M^2}.
\label{app.prop.1}
\end{align}

The diagonalization in the gauge sector is obtained by redefining the Nakanishi-Lautrup multiplier field 
\begin{align}
    b'= b - \xi \partial A - ev \chi .
\end{align}
Then, the $A_\mu$-propagator is
\begin{align}
    \Delta_{\mu\nu} = -i \Big ( \frac{1}{p^2-M_A^2} T_{\mu\nu} + \frac{1}{\xi p^2-M_A^2} \Big ); \qquad M_A = ev,
\end{align}
whereas the the Nakanishi-Lautrup, pseudo-Goldstone and ghost propagators are
\begin{align}
    &\Delta_{b'b'} = i \xi;& &\Delta_{\chi\chi} = \frac{i}{p^2-\frac{M_A}{\xi}};&
    &\Delta_{\bar \omega \omega} = \frac{i}{p^2-\frac{M^2_A}{\xi}}.
\end{align}

Finally, The ghost associated to the constraint BRST symmetry is free:
\begin{align}
    \Delta_{\bar c c} = \frac{-     i}{p^2-m^2}.
\end{align}

\section{Identities for connected amplitudes from the $X_{1}$-equation}\label{app:conn_amps}

Let us check the identities arising from~\1eq{insertion.barc} in the external $T_1,\bar c^*$ sector. In order to obtain the connected amplitudes we need to carry out the Legende transform by substituting $\Phi$ as a function of $J$ by inverting the condition $\Phi = \frac{\delta W}{\delta J}$. We also go on-shell with the external sources $J$, so we need to solve the following equations of motion (we only display the terms required for the evaluation of $W_{\bar c^*}, W_{T \bar c^*}, W_{\bar c^* \bar c^*}$ and move to the momentum space):
\begin{align}
    & (-p^2 + m^2)X_1 + v \bar c^* + \dots = 0  \Rightarrow  X_1(p) = \frac{v}{p^2-m^2} \bar c^*(p), \\
    & -(-p^2 + m^2)\sigma' + v \bar c^* + \dots = 0 \Rightarrow  \sigma'(p) = -\frac{v}{p^2-m^2} \bar c^*(p) .
\label{Xs.connected.subst}
\end{align}
Notice that since at the linearized level $\bar c^*$ in the classical action is coupled as $\sim v \bar c^* (  \sigma -  X_2) =
v \bar c^* (  \sigma' +  X_1) $ there are no contributions from connected diagrams exchanging a tree-level $X_2$-propagator.

\begin{figure}
    \includegraphics[scale=.75]{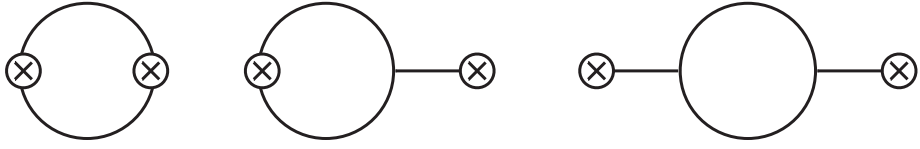}
    \caption{Contributions to $W_{\bar c^* \bar c^*}$ at one loop. Vertices indicate the insertion of $\bar c^*$, whereas straight lines denote the $X_1$ and/or $\sigma'$ propagators.}
    \label{fig.3}
\end{figure}

We list the connected amplitudes to be evaluated:
\begin{itemize}
    \item $W_{\bar c^*} = 0$
    
    At one loop order this amplitude has a 1-PI contribution from $\G^{(1)}_{\bar c^*}$ and a piece associated with the tadpoles of $X_1, \sigma'$ via the replacements in Eq.(\ref{Xs.connected.subst}):
    \begin{align}
    W_{\bar c^*(0)} = \G^{(1)}_{\bar c^*(0)} - \frac{v}{m^2} ( \G^{(1)}_{X_1(0)} -  \G^{(1)}_{\sigma'(0)}) \, .
    \label{W.barc}
    \end{align}
    Notice that since only tadpoles are involved the momentum is set to zero. \1eq{W.barc} can be easily verified.
    
    \item $W_{T_1 \bar c^*} = 0$
    
    In much the same way the connected amplitude $W_{T_1 \bar c^*}$ contains a 1-PI contribution plus a connected piece from $\G^{(1)}_{T_1 X_1}$, $\G^{(1)}_{T_1 \sigma'}$:
    \begin{align}
    W_{T_1 \bar c^*} = \G^{(1)}_{T_1 \bar c^*} + \frac{v}{p^2-m^2} ( \G^{(1)}_{T_1 X_1} -  \G^{(1)}_{T_1 \sigma'}) .
    \label{W.barcT1}
    \end{align}
    Again~\1eq{W.barcT1} can be immediately verified.
    
    \item $W_{\bar c^* \bar c^*} = 0$
    
    The check of this relation is somehow more involved. By collecting all contributions from the diagrams depicted in Fig.~\ref{fig.3}, we obtain
    \begin{align}
        W_{\bar c^* \bar c^*} = \G^{(1)}_{\bar c^* \bar c^*} + \frac{2v}{p^2-m^2}(\G^{(1)}_{\bar c^* X_1} -  \G^{(1)}_{\bar c^* \sigma'}) + \frac{v^2}{(p^2-m^2)^2} ( \G^{(1)}_{\sigma'\sigma'} + \G^{(1)}_{X_1X_1}- 2 \G^{(1)}_{X_1\sigma'}) .
        \label{W.barcbarc}
    \end{align}
    One can check that the sum on the right-hand side does indeed give zero, in agreement with the connected $X_1$-equation.
\end{itemize}
    
\section{One-loop UV divergent contributions to ${\widetilde\G}^{(1)}_{A_\mu A_\nu \sigma}$}\label{app:F2sigma}

We collect here the contributions to ${\widetilde{\overline{\G}}}^{(1)}_{A_\mu A_\nu \sigma}$. After integration by parts in order to leave the $\sigma$ field undifferentiated, a direct calculation yields
\begin{align}
    \int \frac{1}{2} {\widetilde{\overline{\G}}}^{(1)}_{A_\mu A_\nu \sigma} A^\mu A^\nu \sigma = \int \!\mathrm{d}^4 x \Big ( 
    \frac{r_1}{2} \partial^\mu A^\nu \partial_\mu A_\nu 
    + 
    \frac{r_2}{2} A^2  + \frac{r_3}{2} A_\mu \partial^\mu \partial^\nu A_\nu  + \frac{r_4}{2} (\partial A)^2  + \frac{r_5}{2} A_\mu \square A^\mu \Big ) \sigma,
\end{align}
where the UV-divergent coefficients are
\begin{subequations}
	\begin{align}
    	r_1 &= -\frac{1}{4 \pi^2} \frac{g^2 M_A^2}{\Lambda^2 v }\frac{1}{4-D} \, , \label{r1} \\
    	r_2 &= - \frac{M_A^2}{32 v^3 \pi^2} \Big [
    -\frac{g v}{\Lambda}  \Big (4 - 16 \frac{g v}{\Lambda} + 3 \frac{g^2  v^2}{\Lambda^2}\Big ) M^2  + \nonumber \\
    &+
     \Big (-24 - 52 \frac{g v}{\Lambda} - 18 \frac{g^2 v^2}{\Lambda^2} + 3 \frac{g^3 v^3}{\Lambda^3} \Big ) M_A^2
    \Big ] \frac{1}{4-D}, \\
    r_3 &= - \frac{M_A^2}{48 v^2\pi^2}\frac{g}{\Lambda}
    \Big [ 4 + 12 \frac{g v}{\Lambda} - \frac{g^2 v^2}{\Lambda^2} \Big ] \frac{1}{4-D}, \nonumber \\
    r_4 &= - \frac{M_A^2}{16 \pi^2} \frac{g^2}{\Lambda^2 v} \Big ( 2 + \frac{g v}{\Lambda} \Big )  \frac{1}{4-D}, \\
    r_5 &= \frac{M_A^2}{48 v^2\pi^2}\frac{g}{\Lambda}
    \Big [ 4 - 12 \frac{g v}{\Lambda} - \frac{g^2 v^2}{\Lambda^2} \Big ] \frac{1}{4-D}. 
    \label{uv.coeff}
\end{align}
\end{subequations}

\section{\label{app:Gauss}Off-shell equivalence between the $X$ and the target ($\phi$) theory}

The proof of the off-shell equivalence between the $X$ and the target ($\phi$) theory is relatively straightforward, as it only requires a simple saddle point expansion involving only the $X_2$ field. In the case addressed in this paper, this can be even done in a closed form as the $X$ fields enters at most quadratically.

Let us first assume that there is no dimension 6 operator in the classical action~\noeq{cl.g.i.act}. Then, its quadratic part in the $X$ fields can be rewritten as
(in momentum space)
\begin{align}
	S_{X^2}&=-\frac12\sum_{i,j}
	\int d^4 p \, X_iA_{ij}X_j+\sum_i \int d^4p \, X_iJ_i,
\end{align}
where 
\begin{align}
	A&=\left[
\begin{matrix}
    0  & -p^2+m^2 \\
    -p^2+m^2 & -2p^2+M^2+m^2
\end{matrix}
\right];&
J_1=J_2&=\frac1v(-p^2 + m^2) \Big ( \phi^\dagger \phi - \frac{v^2}{2}\Big).
\end{align}

Integration over the fields $X_{1,2}$ yields
\begin{align}
	\int\!{\cal D}X_i\,\mathrm{exp}\left[S_{X^2}\right]=\frac1{\cal N}\mathrm{exp}\left[\frac12 \sum_{i,j}\int J_iA^{-1}_{ij}J_j\right],
\end{align}
where ${\cal N}$ is a source independent normalization factor whereas
\begin{align}
		A^{-1}&=\left[
\begin{matrix}
    \frac{2p^2-M^2-m^2}{(p^2-m^2)^2}  & -\frac1{p^2-m^2} \\
    -\frac1{p^2-m^2} & 0
\end{matrix}
\right].
\end{align}
Thus one obtains
\begin{align}
	\frac12\sum_{i,j}J_iA^{-1}_{ij}J_j=-\frac{M^2-m^2}{2v^2}\Big ( \phi^\dagger \phi - \frac{v^2}{2}\Big)^2,
	\label{intout}
\end{align}
which once summed with the quartic term already present in~\1eq{cl.g.i.act}, gives the usual quartic potential term depending only from the physical mass $M$.

Next, in the presence of the dimension 6 operator $\frac{g}{\Lambda} X_2 (D^\mu \phi)^\dagger (D_\mu \phi)$, the source $J_2$ receives an additional contribution $\delta J_2$ with
\begin{align}
	\delta J_2&=\frac{g}{\Lambda} (D^\mu \phi)^\dagger (D_\mu \phi).
\end{align}  
Then~\noeq{intout} receives the extra term
\begin{align}
	-J_1\frac1{p^2-m^2}\delta J_2=\frac1v\frac{g}{\Lambda}\Big( \phi^\dagger \phi - \frac{v^2}{2}\Big)(D^\mu \phi)^\dagger (D_\mu \phi),
\end{align} 
which is the original operator we wish to describe.

It is clear that the analysis straightforwardly extends to the saddle point approximation in the presence of a non-trivial interaction depending on $X_2$. 
The replacement of the field $\sigma$ in favour of the gauge-invariant variable $X_2$ can then
be consistently carried out, being a matter of convenience (in terms of
simplicity of the defining functional equations as well as the fulfillment of the
power-counting) to choose the most appropriate representation for the higher dimensional operators.


\end{document}